\documentclass[%
reprint, 
10pt,
superscriptaddress,
 amsmath,amssymb, aps,
pra,
longbibliography, 
]{revtex4-2}

\usepackage{amsmath}
\usepackage{graphicx,xcolor}
\usepackage{dcolumn}
\usepackage{bm}
\usepackage[english]{babel}
\usepackage{url}

\usepackage{changes}

\usepackage[normalem]{ulem}

\begin{document}

\preprint{APS/123-QED}

\title{Demonstration of a Tunable Non-Hermitian Nonlinear Microwave Dimer}

\author{Juan S. Salcedo-Gallo}
\affiliation{\mbox{Thayer School of Engineering, Dartmouth College, 15 Thayer Drive, Hanover, New Hampshire 03755, USA}}

\author{Michiel Burgelman}
\affiliation{\mbox{Department of Physics and Astronomy, Dartmouth College, 6127 Wilder Laboratory, Hanover, New Hampshire 03755, USA}}

\author{Vincent P. Flynn}
\affiliation{\mbox{Department of Physics and Astronomy, Dartmouth College, 6127 Wilder Laboratory, Hanover, New Hampshire 03755, USA}}

\author{Alexander S. Carney}
\affiliation{\mbox{Thayer School of Engineering, Dartmouth College, 15 Thayer Drive, Hanover, New Hampshire 03755, USA}}

\author{Majd Hamdan}
\affiliation{\mbox{Thayer School of Engineering, Dartmouth College, 15 Thayer Drive, Hanover, New Hampshire 03755, USA}}

\author{Tunmay Gerg}
\affiliation{\mbox{Department of Physics and Astronomy, Dartmouth College, 6127 Wilder Laboratory, Hanover, New Hampshire 03755, USA}}

\author{Daniel C. Smallwood}
\affiliation{\mbox{Thayer School of Engineering, Dartmouth College, 15 Thayer Drive, Hanover, New Hampshire 03755, USA}}

\author{Lorenza Viola}
\affiliation{\mbox{Department of Physics and Astronomy, Dartmouth College, 6127 Wilder Laboratory, Hanover, New Hampshire 03755, USA}}

\author{Mattias Fitzpatrick}
\email{mattias.w.fitzpatrick@dartmouth.edu}
\affiliation{\mbox{Thayer School of Engineering, Dartmouth College, 15 Thayer Drive, Hanover, New Hampshire 03755, USA}}
\affiliation{\mbox{Department of Physics and Astronomy, Dartmouth College, 6127 Wilder Laboratory, Hanover, New Hampshire 03755, USA}}

\begin{abstract}
Achieving and controlling non-reciprocity in engineered photonic structures is of fundamental interest in science and engineering. Here, we introduce a tunable, non-Hermitian, nonlinear microwave dimer designed to precisely implement {\em phase-non-reciprocal} hopping dynamics between two spatially separated cavities at room temperature. Our system incorporates simple components such as three-dimensional microwave cavities, unidirectional amplifiers, digital attenuators, and a digital phase shifter. By dividing the energy transfer into forward and backward paths, our platform enables precise control over the amplitude and phase of the propagating signals in each direction. Through a combination of theoretical and numerical analysis, we model the dynamics of the system under different operating conditions, including a parameter regime where the gain not only compensates for but significantly exceeds the inherent loss. Our model quantitatively reproduces the observed weak-drive transmission spectra, the amplitude and frequency of self-sustained limit cycles, and the phase locking synchronization effect between the limit cycle and an external microwave tone. Our results may have implications in areas ranging from sensing and synthetic photonic materials to neuromorphic computing and quantum networks, while providing new insight into the interplay between non-Hermitian and nonlinear dynamics. 
\end{abstract}


\maketitle

\section*{\label{sec:intro} 
Introduction}

An isolated quantum system undergoes unitary dynamics, generated by a Hermitian Hamiltonian. Since no system can be perfectly isolated from its surrounding environment, however, non-Hermiticity appears naturally in describing the non-unitary, irreversible evolutions that real-world {\em open} systems undergo \cite{Ashida}. Non-Hermitian effective Hamiltonians have long been used to model a variety of open-system behavior phenomenologically, from the decay of unstable states to anomalous wave propagation and localization \cite{Feshbach, HN}, to gain-and-loss phenomena \cite{El-Ganainy07, Graefe_2010, el-ganainy_non-hermitian_2018} and parity-time (${\mathcal{PT}}$) symmetry-breaking transitions \cite{Bender_1998, Bender_2007, Salamo, bittner_p_2012, Feng2017, Jin2024}. Within a more rigorous treatment in the framework of open quantum systems \cite{Breuer2009}, a broad class of systems undergoing Markovian dissipation may be accurately described by a Lindblad master equation \cite{Lindblad1976}. A probability-non-conserving evolution described by a non-Hermitian effective Hamiltonian then arises in a semiclassical or a measurement-post-selected regime, where quantum fluctuations and quantum jumps can be neglected. Remarkably, as a sole consequence of quantum statistics, effectively non-Hermitian dynamics may also arise for closed systems of non-interacting bosons, despite their Hamiltonian remaining Hermitian at the many-body level \cite{blazoit86, ClerkBKC, Decon, Mariam}. 

Systems evolving under explicitly or even effectively non-Hermitian dynamics can exhibit a wealth of distinctive features, which are both of fundamental interest and can be harnessed for practical applications. Notably, non-Hermiticity makes it possible 
for a system to sustain non-reciprocal couplings, which offers new opportunities for realizing unidirectional, phase-dependent transport and amplification \cite{metelmann_nonreciprocal_2015, Painter, ClerkBKC}, and may be ultimately traced back to the non-trivial topology of the underlying dynamical generator \cite{PorrasTopoAmp,Wanjura2020}. Likewise, enhanced sensing modalities are being predicted to stem from both non-reciprocity \cite{mcdonald_exponentially-enhanced_2020} and uniquely non-Hermitian spectral singularities associated with loss of diagonalizability and the emergence of exceptional points \cite{Wiersig}. Renewed interest in non-Hermitian dynamics is driven by the realization that, in connection with ideas from many-body and topological physics, non-reciprocal interactions may underpin a range of emergent phenomena of broad relevance to photonics, optics, acoustics, and condensed-matter physics \cite{Ashida, CarusottoRMP, Flebus}. Representative examples include novel time-dependent \cite{Vitelli} or transient metastable phases \cite{flynn_topology_2021, flynn_2023}, as well as exotic strongly-correlated phases of matter with non-Hermitian topology \cite{OkumaSato} or broken time-translation symmetry \cite{Liu2023, Raskatla2024}. 

By now, numerous experimental platforms have successfully demonstrated non-Hermitian dynamics and the resulting non-reciprocal transmission in the classical regime. In particular, optomechanical oscillators coupled in a non-reciprocal, nonlinear regime have revealed $\mathcal{PT}$-symmetry-breaking phase transitions and limit-cycle oscillations \cite{brzobohaty_synchronization_2023, reisenbauer_non-Hermitian_2024, liska_pt-like_2024}, and the use of a parametrically driven nano-optomechanical network has led to initial implementations of a paradigmatic (effectively) non-Hermitian model described by a bosonic Kitaev chain Hamiltonian \cite{BKCExp}. Meanwhile, real-world acoustic experiments have demonstrated loss-compensated, non-reciprocal scattering and self-oscillations as well \cite{Pedergnana_2024}.  Within circuit quantum electrodynamics (QED) setups, existing approaches for breaking reciprocity often rely on active cavities \cite{rao_meterscale_2023} or complex schemes that require magnetic-field-tunable components like Yttrium-Iron-Garnet (YIG) spheres \cite{owens_quarter-flux_2018, owens_chiral_2022}. Circuit QED platforms have proven instrumental for realizing exotic topological lattices with complex connectivities and non-Euclidean geometries \cite{Nunnenkamp_2011, Houck2012, kollar_hyperbolic_2019, Carusotto2020}. While they are thus ideally positioned for studies of strongly correlated photonic materials and synthetic gauge fields, the above-mentioned schemes lack the degree of flexibility and tunability that would be desirable for explicitly engineering non-Hermiticity and non-reciprocal hopping dynamics in these systems.

With the above challenge in mind, in this work we demonstrate a highly tunable non-Hermitian device as a fundamental building block toward realizing scalable synthetic photonic lattices. We construct this building block using two coupled three-dimensional (3D) aluminum cavities acting as classical harmonic oscillators at room temperature, and forming a dimer system. Our key idea for engineering non-reciprocal interactions in a way that can be precisely calibrated and controlled is to combine a novel realization of non-Hermitian hopping dynamics, obtained via the insertion of a digital phase shifter, with the intrinsic nonlinearities stemming from amplifier saturation. This effectively results in a closed-feedback network of active and passive elements, which permits access to a regime where gain surpasses inherent loss, forcing the onset of dynamical instability at the linear level. Under these conditions, the system undergoes a supercritical Hopf bifurcation \cite{Holmes}, and the nonlinear dynamics eventually result in a self-sustained, stable limit cycle (LC). 

Besides characterizing the stability phase diagram in the undriven case, we explore the effects of frequency entrainment \cite{Hayashi1964} (a form of synchronization) that the LC undergoes in the presence of an external microwave tone. Phase-locking \cite{Balanov2009} and synchronization phenomena have long been recognized as hallmark features in understanding the classical-to-quantum transition of various nonlinear oscillators \cite{Sadeghpour2013, Walter2014, Walter2015, Zhang_2024, Solanski2025}, 
and they show significant promise for device applications, particularly in injection-locking techniques \cite{backscatter_2020}. Here, we systematically investigate the interplay of gain, loss, non-reciprocity, and nonlinear saturation through experiments, and quantitatively account for all the observed effects using numerical simulations and analytical calculations based on a proposed phenomenological model.  


\section*{\label{sec:results} 
Results}

\begin{figure}
\begin{centering}
\includegraphics[width=0.45\textwidth]{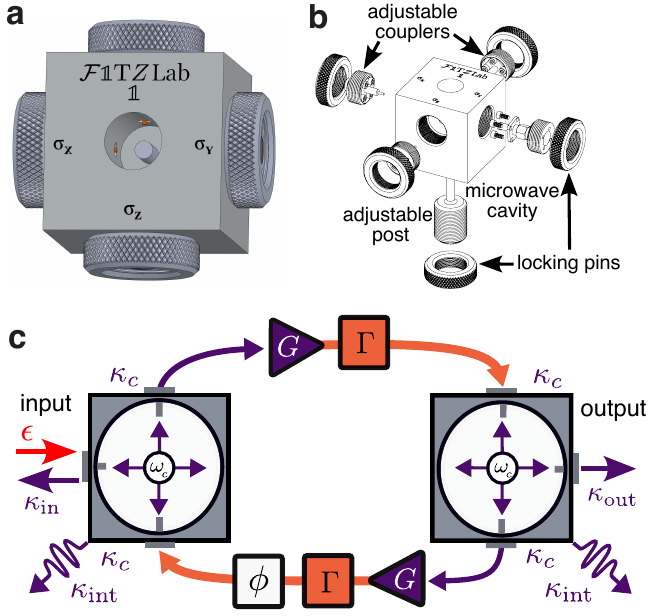}
\end{centering}
\caption{\textbf{Tunable, non-Hermitian, nonlinear microwave dimer.} \textbf{a}, Depiction of individual cavities with a mechanically-adjustable resonance frequency and coupling rates. \textbf{b,} An exploded view of the cavity, showing how cavities are assembled with locking pins to set experimental parameters after calibration. \textbf{c}, Schematic of the dimer system formed from two tunable microwave harmonic oscillators with frequencies $\omega_c/2\pi = 6.027$(5) GHz and internal quality factors of $Q_{\text{int}}= \omega_c/\kappa_{\text{int}} \approx 1488$. The oscillators are connected with a characteristic coupling strength of $\kappa_c / 2\pi = 8.7(1)$ MHz. When the amplifier is in its normal operating regime, this coupling is determined by a characteristic gain, $G_0 = 20.3(2)$ dB, and subsequently adjusted through digital attenuation, $\Gamma$.  On the return path (cavity $2\rightarrow$ cavity $1$), a phase shifter introduces a relative phase, $\phi$, between the two coupling paths. We couple photons into and out of the cavity at an average rate $\kappa_{\text{in, out}} / 2\pi = 4.0(2)$ MHz.}
\label{fig:setup}
\end{figure}

Figure \ref{fig:setup}\textbf{a} illustrates the 3D microwave cavities used in our system, each with four ports featuring adjustable coupling rates, tunable (identical) resonance frequency, $\omega_c$, and fixed internal loss rates, $\kappa_{\text{int}}$. Fig.\,\ref{fig:setup}\textbf{b} presents an exploded view of the cavities, showing the length-adjustable couplers, each formed by the center conductor of a coaxial connector that controls the input, output, and coupling rates, labeled as $\kappa_{\text{in}}$, $\kappa_{\text{out}}$, and $\kappa_c$, respectively. The resonance frequency is tuned by adjusting a post at the base of the device, with locking pins ensuring the stability of all mechanically-tunable parameters after calibration.

As shown in Fig.\,\ref{fig:setup}\textbf{c}, cavity 1 is driven by a coherent signal with strength $\epsilon$ and frequency $\omega_d$ at an input rate $\kappa_{\text{in}}$. Hence, we define $\epsilon \equiv \sqrt{\kappa_{\text{in}}P_{\text{in}}/\hbar\omega_d}$, where $P_{\text{in}}$ is the input drive power (in watts), converted from the corresponding value $P_d$ (in dBm) using $P_{\text{in}} = 10^{(P_d - 30)/10}$.

Both cavities are coupled to unidirectional amplifiers with a characteristic gain $G_0$, followed by a digital attenuator with a dynamic range of $\Gamma \in [0, 50]$ dB, adding to the intrinsic insertion loss of the passive components. A digital phase shifter is inserted to make the relative phase, $\phi$, of the reverse propagating path (cavity 2 $\to$ cavity 1) {\em tunable} over $[0, 2\pi)$, and the output is collected from cavity 2 at a rate $\kappa_{\text{out}}$.  In what follows, we characterize our system using the quantity $\Delta G \equiv G_0 - \Gamma$, representing the {\em net} hopping gain. By utilizing a conversion factor of $10^{\Delta G/20}$, we adjust the intrinsic $\kappa_c$ of each oscillator to produce an effective, {\em tunable} hopping coefficient at low power given by $J_0(\Delta G) = 10^{\Delta G/20} \kappa_c$. 

Before describing the full nonlinear model of our system, we present a linear model that accurately describes the essential features of the dynamics in the low-power regime. This is achieved by considering the following semiclassical equations of motion (EOMs):
\begin{eqnarray}
\label{eq:dynamics}
\dot{\bm{\alpha}} &=& A_{0}
\bm{\alpha} + \epsilon\textbf{B},  
\end{eqnarray}
where the state variables $\bm{\alpha}\equiv \begin{bmatrix}
\alpha_1 \: \alpha_2
\end{bmatrix}^T$ represent complex amplitudes of the cavity field, 
$\textbf{B} 
\equiv \begin{bmatrix} 1 \:0
\end{bmatrix}^T$ accounts for the external drive on cavity 1, and the dynamical matrix $A_0$ takes the form
\begin{align}
\label{eq:general_form}
A_0=  \begin{bmatrix}
-i(\omega_c - \omega_d) - \kappa_0 & -iJ_0(\Delta G) e^{-i\phi} \\ -iJ_0(\Delta G) & -i(\omega_c -\omega_d) - \kappa_0
\end{bmatrix}.
\end{align}
In this way, tuning the phase $\phi$ takes the system from a regime where the hopping dynamics is perfectly reciprocal ($\phi=0$), to one where the couplings are skew-Hermitian  ($\phi=\pi$). Thus, {\em phase non-reciprocity} occurs in our system when the relative phase of the hopping terms differ, despite the hopping rates remaining equal. As a key difference from existing non-reciprocal devices, our dynamical matrix $A_0$ is non-Hermitian but nonetheless {\em normal} (hence diagonalizable) throughout the undriven parameter regime.

In Eq.\,\eqref{eq:general_form}, we have introduced the parameter $\kappa_0$ to describe the total intra-cavity dissipation rate. Experimental observations suggest that $\kappa_0$ is also influenced by $J_0(\Delta G)$. This influence can be explained by the increase in state amplitudes within the cavities, $\alpha_i$, which leads to the amplifiers introducing and amplifying existing incoherent noise over a finite bandwidth determined by the inherent linewidth of the cavities. Since this amplified noise competes directly with the intrinsic dissipation of each cavity, the overall effective dissipation rate is reduced. We model this process phenomenologically as:
\begin{align}
\label{eq:diss_rate_deltaG}
\kappa_{0} \equiv \kappa_0(\Delta G) = 2\,(\kappa_{\text{int}} + \kappa_{\text{in/out}} + \kappa_c) - J_0(\Delta G).
\end{align}

\begin{figure}[htbp!]
\includegraphics[width=0.5\textwidth]{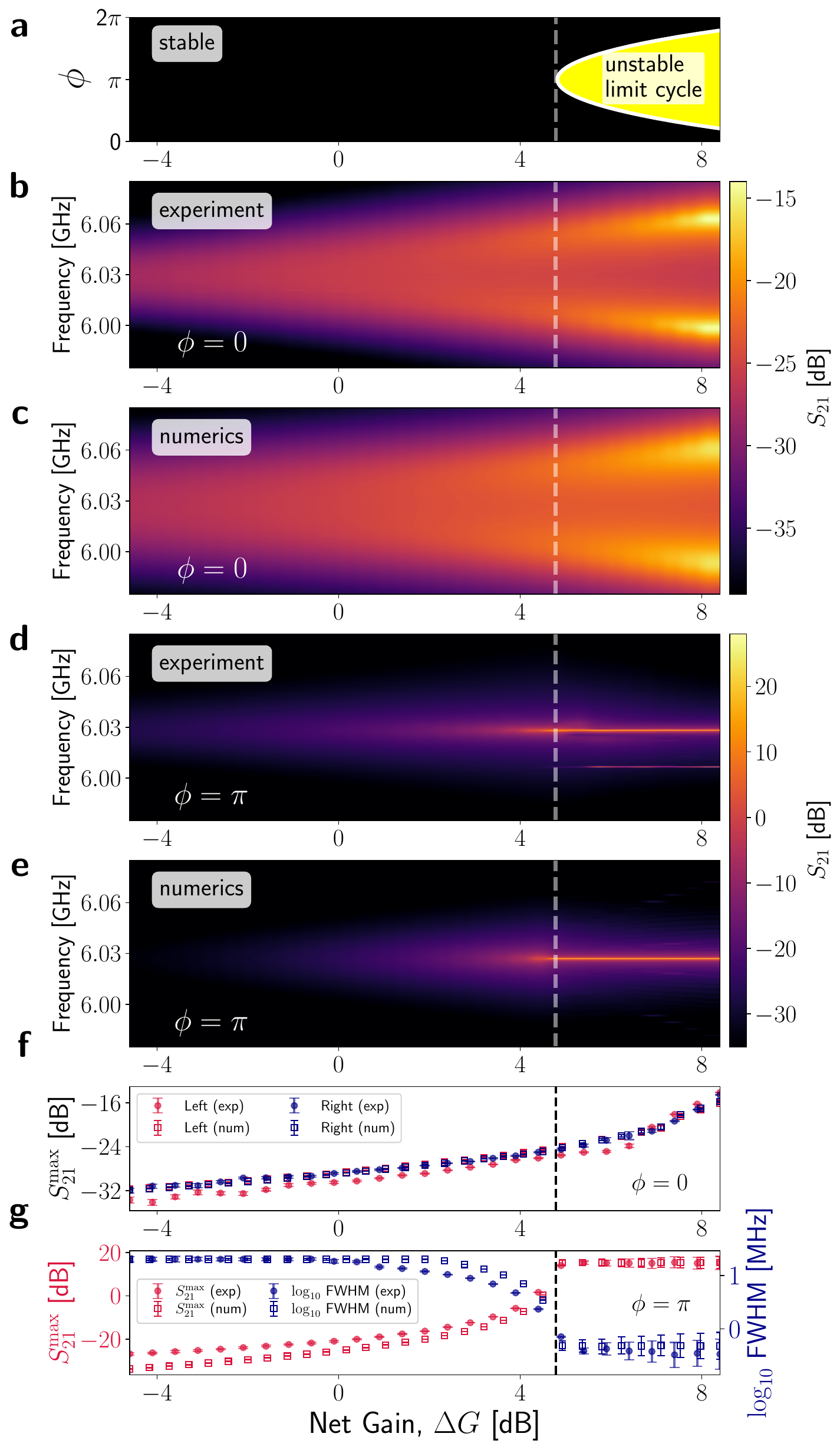} \caption{\textbf{Stability phase diagram and transmission spectra $\bm{(S_{21})}$ for our tunable non-Hermitian nonlinear microwave dimer}. \textbf{a}, Stability phase diagram depicting the vacuum-stable and unstable regimes determined by the dynamical matrix in Eq.\,\eqref{eq:general_form} or Eq.\,\eqref{eq:model-full}. \textbf{b(d)}, Experimental, and \textbf{c(e)} numerical simulations for $S_{21}$ as a function of $\Delta G$ for $\phi = 0(\pi)$. In both cases, the data are plotted with the same colorbar to demonstrate the agreement between experimental and numerical results. Numerical simulations (squares) and experimental measurements (circles) of maximum transmission, $S_{21}^{\text{max}}$, are shown in \textbf{f–g}. A comparison of the experimental full width at half maximum (FWHM) with corresponding numerical results for $\phi = \pi$ is included in \textbf{g}, showing a dramatic reduction in linewidth as $\Delta G$ increases. $S_{21}^{\text{max}}$ and FWHM are extracted from single- or double-Lorentzian fits to the spectra in \textbf{b–c} and \textbf{d–e}, with FWHM plotted on a logarithmic scale to emphasize the sharp transition to sub-MHz linewidhts above threshold. Experimental and numerical results in \textbf{b–g} consider an input drive power of $P_{d} = -30$ dBm. The dashed line at $\Delta G = 4.78$ in \textbf{a–g} denotes the onset of instability at $\phi = \pi$.}
\label{fig:fig_2}
\end{figure}

Generally, Eq.\,\eqref{eq:dynamics} admits a unique stable equilibrium point when $\max \text{Re} [\sigma(A_0)] < 0$, where $\sigma(A_0)\equiv \sigma(A_0(\Delta G, \phi))$ denotes the eigenvalue spectrum of $A_0$ as a function of the tunable parameters. Thus, a necessary and sufficient condition for dynamical stability is that 
\begin{align}
\label{eq:linstabeq}
\frac{J_0(\Delta G)}{\kappa_0(\Delta G)} \sin(\phi/2) < 1 .
\end{align} 
The associated stability phase boundary as a function of $\phi$ and $\Delta G$ is illustrated in Fig.\,\ref{fig:fig_2}\textbf{a}. From Eq.\,\eqref{eq:linstabeq}, we can immediately see that for $\phi \neq 0$ and sufficiently high $\Delta G$, it is possible that $J_0(\Delta G)$ can balance and exceed $\kappa_0(\Delta G)$, resulting in the onset of instability. This condition selects two relevant regions, namely,
\begin{align}
\text{Region I: }& 0 \leq J_0(\Delta G) \leq \kappa_0(\Delta G) ,
\\
\text{Region II: }& \kappa_0(\Delta G) < J_0(\Delta G).
\end{align} 
Region I is always stable, while Region II is only stable for those $\phi$ satisfying Eq.\,\eqref{eq:linstabeq}. Physically, in the gain-dominated (Region II) unstable regime, the field amplitudes $|\alpha_i|^2$ diverge, pushing the amplifiers into saturation and reducing the gain in a power-dependent fashion.

While the linear model captures the correct asymptotic behavior in the stable, loss-dominated (Region I) regime, to obtain a description of the dynamical behavior valid in both the above regions, we must allow the hopping function to depend nonlinearly upon $|\alpha_i|^2$. We account for such a dependence through the following continuous piecewise function:
\begin{align}
\frac{J (\Delta G; |\alpha_i|^2)}{\kappa_c 10^{\Delta G/20} } &= 
\begin{cases} 
 1 & \text{if } |\alpha_i|^2 \leq |\alpha_\text{sat}|^2, \\
\frac{b_G + \hbar \omega_c |\alpha_{\text{sat}}|^2 \kappa_c}{b_G + \hbar \omega_c |\alpha_i|^2 \kappa_c} & \text{if } |\alpha_i|^2 > |\alpha_\text{sat}|^2 ,
\end{cases}
\label{eq:j_hop}
\end{align}
where $b_G = 8.6$ mW and $|\alpha_{\text{sat}}|^2$ is the saturation threshold of the amplifier, which is determined by $|\alpha_{\text{sat}}|^2 = P_{\text{sat}} / \hbar \omega_c \kappa_c$, with $P_{\text{sat}} = 0.9981$ mW derived from experimental characterization (see Supplementary Information Sec. VI). Note that for sufficiently low $|\alpha_i|^2$, Eq.\,\eqref{eq:j_hop} consistently recovers the hopping coefficient for the linear model, namely $J_0(\Delta G)$. However, as $|\alpha_i|^2$ exceeds $|\alpha_{\text{sat}}|^2$, $J(\Delta G, |\alpha_i|^2)$ becomes monotonically reduced, illustrating the saturation effect that drives the nonlinear behavior of the system.

As we will further discuss in the next section, transmission experiments (Fig.\,\ref{fig:fig_2}\textbf{b,d}) show that peak-splitting at $\phi = 0$ is first resolved at a {\em significantly lower} $\Delta G$ than the one at which instability sets in for $\phi = \pi$. However, Eq.\,\eqref{eq:general_form} indicates that the two normal modes become spectrally resolvable, namely when their splitting equals the peak linewidth, only at $\Delta G \simeq 4.78$ dB, so it predicts that the resolvable splitting and the instability threshold should coincide. This suggests an additional coherent hopping effect that is stronger at $\phi = 0$, allowing for earlier mode splitting (see also Supplementary Information, Sec.\, IB). We make this intuition precise by directly modifying the off-diagonal hopping coefficients via a phase-dependent function, $f(\phi) = i J_c \cos\left(\tfrac{\phi}{2}\right) e^{i\phi/2},$ where $J_c/2\pi = 11.5$ MHz represents the strength of this additional coherent hopping. While a complete explanation of the origin of $f(\phi)$ is lacking, and its proposed $\phi$-dependence is phenomenological, we believe it stems from constructive interference between the two modes. The incorporation of $f(\phi)$ accurately captures the frequency splitting observed in Fig.\,\ref{fig:fig_2}\textbf{b-e} without altering the stability characteristics of the system in Fig.\,\ref{fig:fig_2}\textbf{a}.

\begin{figure}[htbp!]
\centering
\includegraphics[width=0.5\textwidth]{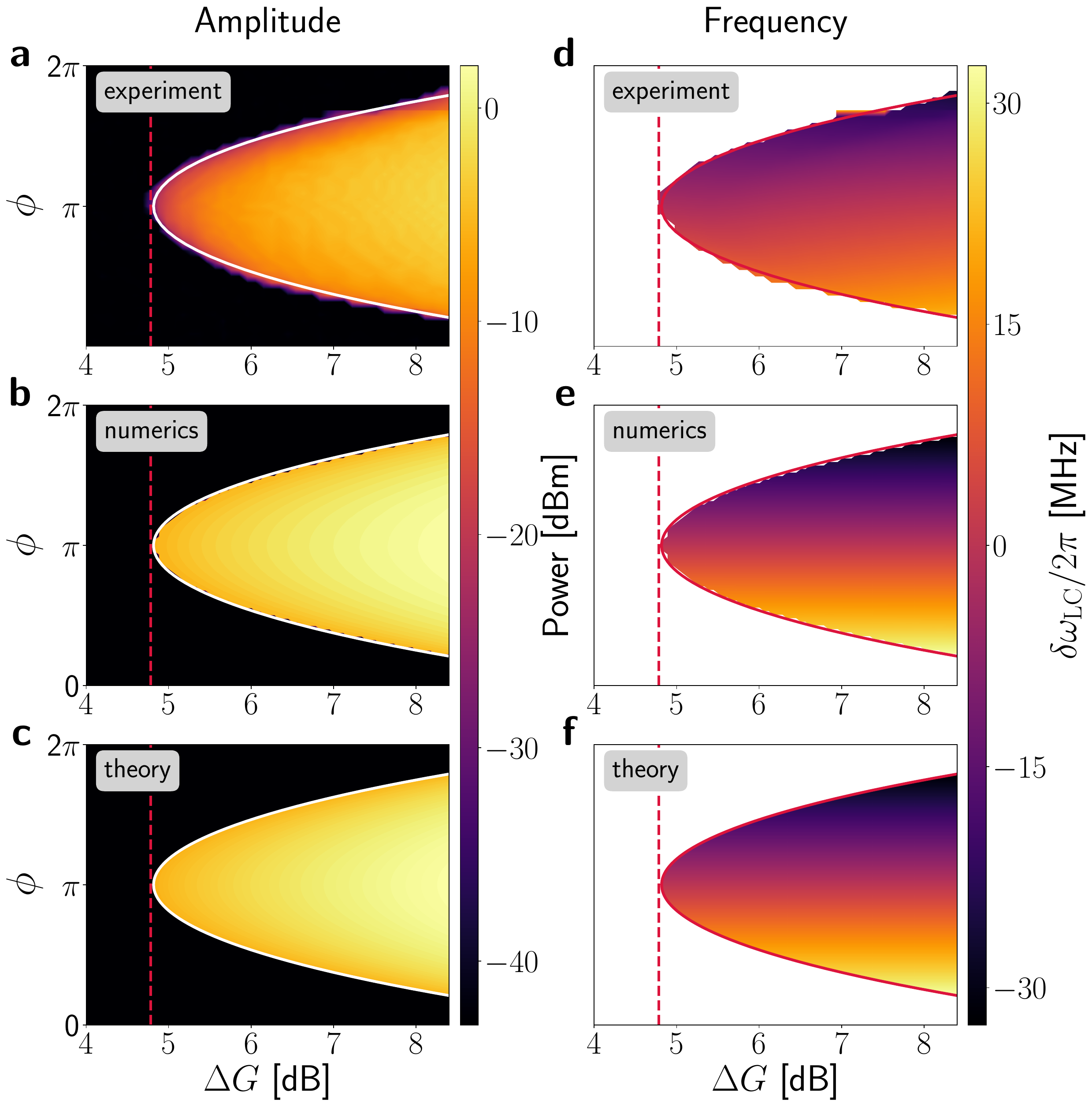}
\caption{\textbf{Phase diagram of the amplitude and frequency of the LC solutions without external driving}.  \textbf{a-c}, Power emitted from the LC extracted experimentally (\textbf{a}), numerically (\textbf{b}), and analytically (\textbf{c}). \textbf{d-f}, Frequency detuning of the LC relative to $\omega_c$, $\delta \omega_{\text{LC}}$, extracted experimentally (\textbf{d}), numerically (\textbf{e}), and analytically (\textbf{f}). The contour in each plot represents the stability phase boundary from the dynamical matrix in Eq.\,\eqref{eq:model-full}, also corresponding to  $|\alpha_2|^2 = |\alpha_{\text{sat}}|^2$ for the analytical solution in Eq.\,\eqref{eq:amp_lc}, which is depicted in \textbf{c, f}. Numerical and analytical amplitudes in the vacuum-stable phase in \textbf{b-c} are set at -44 dBm, corresponding to the baseline amplitude observed experimentally in \textbf{a}, and this same threshold is used to white out the corresponding region in \textbf{d}. The vertical dashed line in all plots marks the value of $\Delta G = 4.78$ dB at which the phase $\phi = \pi$ becomes unstable. In the experimental data, we removed an outlier near $\phi = 5.287$ rads., corresponding to the set value on the digital phase shifter moving from $2\pi\rightarrow 0$.}
\label{fig:fig_3}
\end{figure}

At this point, we have already introduced all of the terms required to define the final form of our EOMs:
\begin{eqnarray}
\label{eq:dynamics_full}
\dot{\bm{\alpha}} &=& A(|\alpha_1|^2, |\alpha_2|^2) \bm{\alpha} + \epsilon\textbf{B},  
\end{eqnarray}
with the full dynamical matrix being given by 
\begin{widetext}
\begin{align}\label{eq:model-full}
A(|\alpha_1|^2, |\alpha_2|^2) &= \begin{bmatrix}
-i(\omega_c - \omega_d) -\kappa_1(\Delta G;|\alpha_1|^2) & [-iJ(\Delta G;|\alpha_2|^2) - f(\phi)] e^{-i\phi} \\ -iJ(\Delta G; |\alpha|_1^2) - f(\phi) & -i(\omega_c-\omega_d) - \kappa_2(\Delta G ;|\alpha_2|^2)
\end{bmatrix}, 
\end{align}
\end{widetext}
\noindent 
where $\kappa_{1(2)} = 2(\kappa_{\text{int,1(2)}} + \kappa_{\text{in(out)}} + \kappa_c) - J(\Delta G, |\alpha_{1(2)}|^2)$, and $J(\Delta G, |\alpha_{1(2)}|^2)$ is defined in Eq.\,\eqref{eq:j_hop}. In the linear limit of small $|\alpha_i|^2$, $A(|\alpha_1|^2, |\alpha_2|^2)$ reduces to $A_0$ as described in Eq.,\eqref{eq:general_form}, up to the phase-dependent correction introduced via $f(\phi)$. By explicitly solving the dynamics defined by Eqs.\,\eqref{eq:dynamics_full} and \eqref{eq:model-full}, we can directly reproduce and capture the main features observed in our experiments, such as weak-drive transmission (Fig.\,\ref{fig:fig_2}\textbf{b-g}), undriven LC solutions (Fig.\,\ref{fig:fig_2}\textbf{a} and Fig.\,\ref{fig:fig_3}), as well as the phase locking synchronization phenomena (Fig.\,\ref{fig:fig_4}). We provide an in-depth discussion in the sections that follow.

\subsection*{\label{sec:driven-transmission} 
Weak-drive transmission spectra}

To probe the steady-state $S_{21}$, we drive the system at an input power of $P_d = -30$ dBm and sweep the drive frequency $\omega_d/2\pi$ over the range $5.98$ to $6.09$ GHz using a scalar network analyzer. Numerically, we compute $S_{21}$ as the ratio of the steady-state output power from cavity 2 and the input drive power in cavity 1. Hence,
\begin{align*}
    S_{21}(\omega_d,\Delta G,\phi) = \frac{P_\text{out}}{P_\text{in}} = \frac{\hbar \omega_d|\alpha_2^\text{eq}|^2 \kappa_\text{out}}{\hbar \omega_d \epsilon^2/\kappa_\text{in}} = \kappa_\text{in}\kappa_\text{out}\frac{|\alpha_2^{\text{eq}}|^2}{\epsilon^2},
\end{align*}
where $\alpha_{2}^{\textrm{eq}}$ is the DC-component of the unique asymptotic solution $\alpha_{2}(t)$. Note that the dependence on $\Delta G$ and $\phi$ is implicitly contained in the solutions for $\alpha_2^\text{eq}$. Moreover, to directly compare numerical solutions to experimental results, the conversion to dB is accomplished via $S_\text{21}$ [dB] $=10\log_\text{10}(S_{21})$.

Figures \ref{fig:fig_2}\textbf{b-g} present the experimental and numerical $S_{21}$ data for the two special phases, $\phi = 0$ and $\phi = \pi$, for which, as noted, the hopping amplitudes are Hermitian and skew-Hermitian, respectively. Specifically, Figs.\,\ref{fig:fig_2}\textbf{b-c} depict the experimental and numerical $S_{21}$ spectra for $\phi = 0$ across various $\Delta G$ values. At low $\Delta G$, the $S_{21}$ spectra show a single, very broad peak, indicating that the two modes are decoupled due to the high effective loss in the hopping path. As $\Delta G$ increases, two distinct peaks emerge and move further apart, demonstrating enhanced coupling and symmetric energy distribution between the two modes. The increasing separation of the peaks and higher $S_{21}$ magnitudes (indicated by brighter colors at higher $\Delta G$ in Figs.\,\ref{fig:fig_2}\textbf{b-c}) confirm the successful experimental implementation of reciprocal hopping with tunable rates, as predicted by the dynamical matrix in Eq.\,\eqref{eq:model-full} at $\phi = 0$. Additionally, the excellent agreement between experimental results and numerical simulations presented in Figs.\,\ref{fig:fig_2}\textbf{b-c} validates the accuracy of our model by capturing the $S_{21}$ characteristics when the coupling coefficients in both paths are engineered to be nominally equal.

Figures \ref{fig:fig_2}\textbf{d-e} show the experimental and numerical $S_{21}$ spectra at $\phi = \pi$. At low $\Delta G$ values, we observe a minimal $S_{21}$ signal, again indicating that the two modes are essentially decoupled. As $\Delta G$ increases to intermediate levels, the signal response increases, but the transmission spectra differ significantly from the typical frequency splitting seen in symmetrically coupled modes. Instead of two peaks in frequency, we observe a {\em single} peak centered at $\omega_c/2\pi$. When $\Delta G\simeq4.78$ dB and beyond, the amplitude of the peak at $\omega_d \simeq \omega_c$ suddenly and dramatically increases, followed by asymptotic saturation, while its linewidth sharply narrows (Fig.\,\ref{fig:fig_2}\textbf{g}), consistent with the onset of a stable LC. \added{}

Since the scalar network analyzer performs homodyne detection at the drive frequency, the experimental $S_{21}$ measurements effectively collect the steady-state transmission amplitude at $\omega_{d}/2\pi$. We account for this effect into our numerical framework to generate the data presented in Fig.\,\ref{fig:fig_2} by breaking down the time-domain signal into its frequency components and extracting the DC component, as described in the Methods section. Overall, the tunable $S_{21}$ behavior highlights a fundamental difference in the system dynamics when the hopping coefficients have equal magnitude but opposite signs, yielding a skew-Hermitian dynamical matrix in Eq.\,\eqref{eq:model-full}, which is realized when $\phi = \pi$.
 
Furthermore, Fig.\,\ref{fig:fig_2}\textbf{f} shows $S_{21}^{\text{max}}$ at $\phi = 0$, where both experimental (solid circles) and numerical (open squares) data display a monotonic increase with $\Delta G$. This trend demonstrates the reduction of intra-cavity dissipation rates in the system as  $\Delta G$ increases, which is captured by Eq.\,\eqref{eq:diss_rate_deltaG}. Similarly, Fig.\,\ref{fig:fig_2}\textbf{g} presents $S_{21}^{\text{max}}$ and FWHM at $\phi = \pi$. Here, near $\Delta G \approx 4.78$ dB, $S_{21}^{\text{max}}$ suddenly increases and saturates, while the FWHM sharply decreases due to amplifier saturation. This behavior indicates that the system has transitioned into a dynamically {\em unstable} regime, which we analyze in detail next.

\subsection*{\label{sec:undriven-limit-cycle} 
Undriven, self-sustained, limit cycles}

So far, we have explored the dynamics of our system under the influence of a (weak) coherent external drive applied to cavity 1, specifically at two phases, $\phi = 0$ and $\phi = \pi$. However, the system hosts a {\em self-sustained} LC in the unstable regime. The stability condition in Eq.\,\eqref{eq:linstabeq}, which is depicted in Fig.\,\ref{fig:fig_2}\textbf{a}, determines the existence of this LC, in the absence of external driving. Figure\,\ref{fig:fig_3} presents the experimental, numerical, and analytical results for the amplitude (\textbf{a-c}) and frequency (\textbf{d-f}) of these LC solutions. Each subplot includes the onset of instability for comparison, with the red dashed line indicating the lowest $\Delta G$ value where instability occurs at $\phi = \pi$.

Figure \ref{fig:fig_3}\textbf{a} presents the experimentally measured amplitude of the LC, obtained from the emission spectra for multiple values of $\phi$ and $\Delta G$ (see Supplementary Information Sec. IIE). In regions where the system is dynamically stable, Fig.\,\ref{fig:fig_3}\textbf{a} shows a nominally constant baseline, indicating the absence of a self-sustained emission signal. However, as the system enters the unstable regime, a distinct peak emerges, seeded by thermal noise at room temperature. The limit cycle arises from a dynamical instability of the zero-amplitude state and does not require external driving. Once the linear stability condition is violated, small fluctuations are amplified by the system’s effective net gain, resulting in a self-sustained steady-state oscillation that is accurately captured without the need to explicitly model noise.
This manifests as a sharp increase in amplitude and an {\em ultra-narrow linewidth}, on the order of kHz. Notably, the transition into the unstable regime is marked by a sudden increase in the intensity of the emitted light (or LC amplitude), indicating that the system has undergone a supercritical Hopf bifurcation \cite{Holmes}. 

We perform numerical simulations to determine the amplitude of the LC solutions, as shown in Figs.\,\ref{fig:fig_3}\textbf{b}. In our numerical simulations, we time-evolve the EOMs with $\epsilon=0$ in Eq.\,\eqref{eq:dynamics_full} and $\omega_d = \omega_c$ in Eq.\,\eqref{eq:model-full}. This allows us to calculate the steady-state intensity of the emitted light from cavity 2 ($\propto |\alpha^{\textrm{eq}}_2|^2$). Additionally, we derive an analytical expression for the LC-amplitude. We do this by transforming the EOMs into the normal mode basis, excluding the eigenvalue corresponding to the stable normal mode and solving for the amplitude $|\alpha_i|^2$ that yields non-trivial solutions. We obtain the following expression for the amplitude of the LC, $n_{\text{LC}}$, (see the Methods section for a summary, and the Supplementary Information, Sec.\, II  for a full derivation):
\begin{widetext}
\begin{equation}
    n_{\textrm{LC}} \equiv {|{\alpha_{1}(t)}}|^2  \equiv {|{\alpha_{2}(t)}}|^2 = \frac{10^{\Delta G / 20} \left(1 + \sin{\left(\frac{\phi}{2} \right)}\right) \left(\kappa_{\textrm{c}}^{2} \hbar \omega_{\textrm{c}}  {|\alpha_{\textrm{sat}}|}^2 + \kappa_{\textrm{c}} b_{\textrm{G}}\right) - 2 b_{\textrm{G}} \left(\kappa_{\textrm{c}} + \kappa_{\textrm{in/out}} + \kappa_{\textrm{int}}\right)}{2 \kappa_{\textrm{c}} \omega_{\textrm{c}} \hbar \left(\kappa_{\textrm{c}} + \kappa_{\textrm{in/out}} + \kappa_{\textrm{int}}\right)}
, \quad  \forall t.
\label{eq:amp_lc}
\end{equation}
\end{widetext}

\begin{figure*}[t]
\begin{centering}
\includegraphics[width=\textwidth]{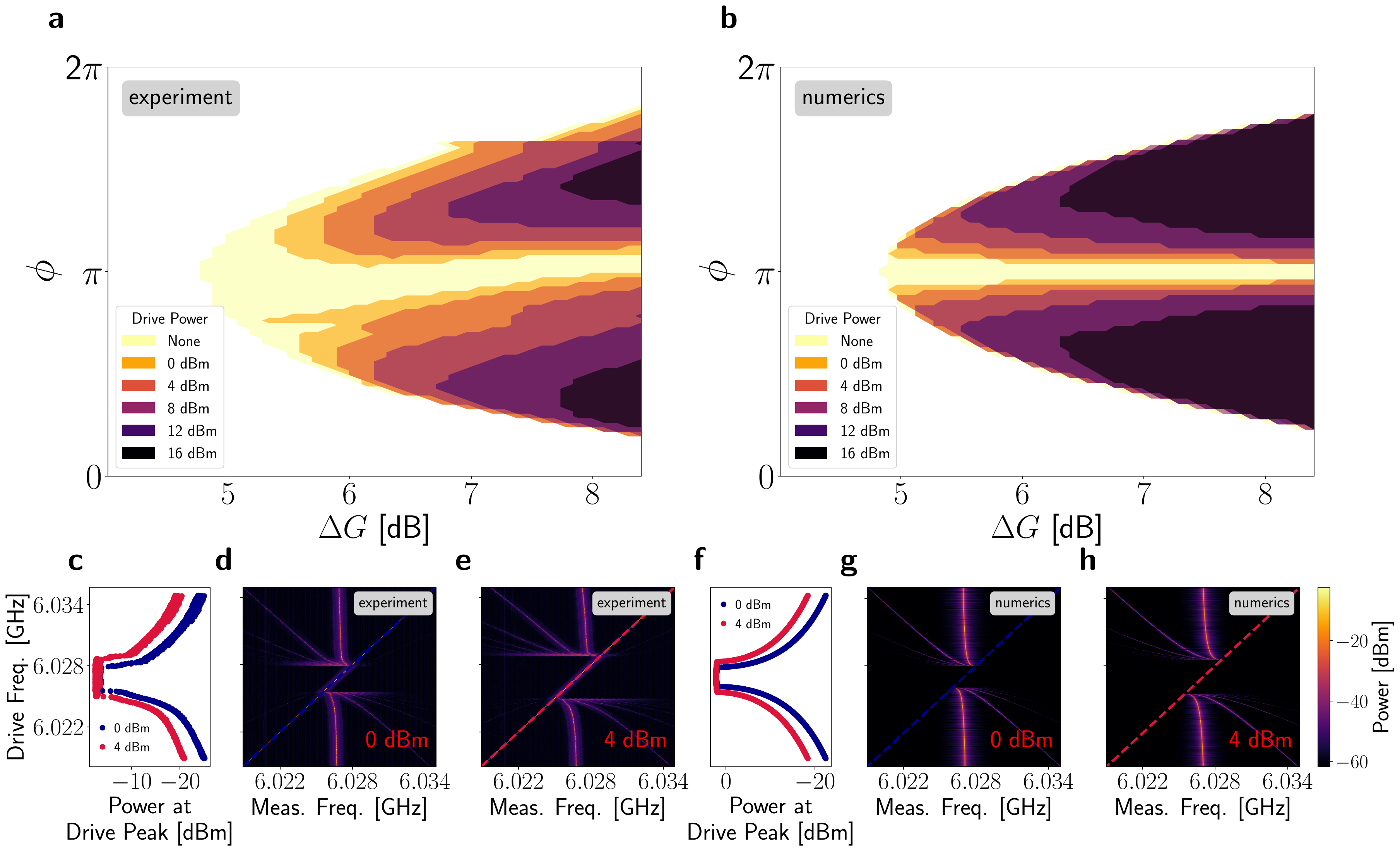}
\end{centering}
\vspace*{-2mm}
\caption{\textbf{Synchronization of the LC mode and an external drive.} \textbf{a-b}, Experimental (\textbf{a}) and numerical (\textbf{b}) contours representing regions containing a distinct LC away from $\omega_c/2\pi$ as a function of drive power. For the experimental and numerical data presented in panels \textbf{a} and \textbf{b}, we applied an external drive at $\omega_d/2\pi = \omega_c/2\pi = 6.027$ GHz. Note that in the LC region of \textbf{a} and \textbf{b}, if $\phi = \pi$, then $\omega_c = \omega_d = \omega_{LC}$, whereas for $\phi \neq \pi$, we have $\omega_c = \omega_d \neq \omega_{LC}$. Experimental (\textbf{c}) and numerical (\textbf{f}) power of the drive peak as a function of the drive frequency for the panels presented in \textbf{d-e} and \textbf{g-h} depicted as blue and red dashed lines for $P_d$ = 0 and 4 dBm, respectively Experimental (\textbf{d-e}) and numerical (\textbf{g-h}) emission from the dimer at $\phi = \pi$ and $\Delta G = 8.4$ dB as an external drive is swept from low to high frequency around $\omega_c/2\pi$ for $P_d = 0$ dBm (\textbf{d,g}), and for $P_d = 4$ dBm (\textbf{e, h}) drive strengths, showing mixing and phase locking synchronization with the external drive. In both experimental and numerical simulations, the LC synchronizes in a larger range in frequency as $P_d$ increases, consistent with the expansion of synchronized regions shown in \textbf{c} and \textbf{f}, which explains the trend observed in the vicinity of $\phi = \pi$ for the contours depicted in \textbf{a-b}, respectively.}

\label{fig:fig_4}
\end{figure*}

In addition to characterizing the amplitude of the self-sustained mode, we also investigate the tunable frequency response. Figure\,\ref{fig:fig_3}\textbf{d-f} displays experimental, numerical, and analytical data that describe the frequency of the LC solutions for a range of $\Delta G$ and $\phi$ values. Specifically, Fig.\,\ref{fig:fig_3}\textbf{d} presents the experimentally measured frequency of the LC, identified as the frequency at which the emission spectra exhibit its highest amplitude (see Supplementary Information Sec. IIE). We calculate the difference $\delta \omega_{\text{LC}}\equiv  \omega_c - \omega_{\text{LC}}$. We see that $\delta \omega_{\text{LC}}$ shifts monotonically with changes in $\phi$, and it is anti-symmetric around $\phi = \pi$. Specifically, $\delta \omega_{\text{LC}}/2\pi$ is positive for $\phi < \pi$ and negative for $\phi > \pi$, thus demonstrating remarkable tunability over a $\sim 60$ MHz frequency range. Importantly, $\delta \omega_{\text{LC}}$ shows negligible dependence on $\Delta G$, which aligns with our numerical and analytic solutions, as discussed next.


To numerically determine $\delta \omega_{\text{LC}}/2\pi$, we perform a Fourier analysis of the time-domain solutions to Eq.\,\eqref{eq:dynamics_full}, and extract the dominant frequency from the spectra, as discussed in the Methods section. Analytically, we derive an expression for $\delta \omega_{\text{LC}}$, revealing that this quantity is independent of $\Delta G$, and entirely determined by $\phi$:
\begin{equation}
\label{eq:LC_frequency}
    \delta \omega_{\textrm{LC}}\equiv J_{c} \cos{\left(\frac{\phi}{2} \right)} + \frac{2 \left(\kappa_{\textrm{in/out}} + \kappa_{\textrm{int}} + \kappa_{c}\right) \cos{\left(\frac{\phi}{2} \right)}}{1 + \sin{\left(\frac{\phi}{2} \right)}}. 
\end{equation} 
Figure \ref{fig:fig_3}\textbf{e-f} present numerical and theoretical calculations of $\delta \omega_{\text{LC}}$ for various $\phi$ and $\Delta G$ values. These results accurately reproduce the key experimental observations, including the monotonic tuning of $\delta \omega_{\text{LC}}$ with $\phi$ over a wide frequency range and a $4\pi$-periodicity. This mechanism enables precise control over the frequency of the self-sustained mode by tuning the relative phase between the hopping paths. Overall, the remarkable agreement between our experimental, numerical, and analytical results demonstrates the effectiveness of our model in explaining the key characteristics of the self-sustained emission process observed in the unstable, gain-dominated regime.

\subsection*{\label{sec:syncrhonization-dynamics} 
Synchronization dynamics}

To further probe the nonlinear dynamics of our device, we now examine a particular synchronization effect, namely, the {\em frequency entrainment} phenomenon or phase locking between the LC and an external microwave tone \cite{Hayashi1964, Balanov2009}. In our experiments and simulations, shown in Fig. \ref{fig:fig_4} \textbf{a-b}, we employ a fixed microwave tone set at $\omega_d = \omega_c$, and vary the drive power, $P_d$, from 0 to 16 dBm. We then analyze the spectra to identify regions of the phase diagram with at least two distinct peaks, indicating the coexistence of self-oscillation and external drive. These regions define the contours in Fig.\,\ref{fig:fig_4}\textbf{a-b}. From here, we observe that increasing the drive power progressively reduces the area where LC solutions exist. This is more pronounced when the LC frequency is close to the drive frequency, which occurs when $\phi = \pi$. Everywhere else, the dynamics converge to a unique stable equilibrium point, where no self-sustained oscillatory behavior occurs. Moreover, we observe a slight asymmetry in the experimental contours of Fig. \ref{fig:fig_4}\textbf{a}, which we attribute to minor experimental imperfections not captured by the model, as discussed further in the Methods section.

To better understand the interaction between the LC mode and the external drive, we perform experiments and numerical simulations sweeping an external tone around $\omega_c/2\pi$ at fixed drive powers (0 and 4 dBm), keeping the dimer set at $\phi = \pi$ and $\Delta G = 8.4$ dB, as shown in Figs.\,\ref{fig:fig_4}\textbf{c-h}. As shown in \textbf{d-e} and \textbf{g-h}, when $\omega_d/2\pi$ is below $\omega_c/2\pi$, the measured spectra show three distinct peaks: the drive-response peak at $\omega_d/2\pi$ (left), a central peak at $\omega_c/2\pi$ from the LC, and a higher harmonic at higher frequencies created by nonlinear wave mixing. As the drive frequency approaches the LC frequency, a pronounced line-pulling effect causes all peaks to coalesce into a single resonance. Moreover, we can see from Figs.\,\ref{fig:fig_4}\textbf{c, f} that in this synchronization window, the power of the drive peak increases and then saturates, further confirming the onset of synchronization. This convergence and line-pulling is a clear signature of a frequency entrainment effect \cite{Hayashi1964}, and phase locking \cite{Balanov2009}, where the LC, with its time-dependence composed of generated higher-order harmonics, coalesces into a single frequency corresponding to that of the external drive. In our setup, phase locking is evidenced by the progressive shift of the dominant spectral peak toward the drive frequency and the merging of harmonics into a single tone, distinct from suppression of natural dynamics, which would result in a stationary peak fading with increasing drive power (see Supplementary Information Sec. III). Furthermore, the synchronization window around $\omega_c/2\pi$ expands with increased drive power, as evident from Figs.\,\ref{fig:fig_4}\textbf{c, f}, hence explaining the observed widening of the gap around $\phi = \pi$ in Figs.\,\ref{fig:fig_4}\textbf{a-b}. At higher drive frequencies, the self-oscillation reappears as a distinct resonance, accompanied by asymmetric higher harmonics, mirroring the behavior observed for drive frequencies below $\omega_c/2\pi$.

\section*{\label{sec:discussion} 
Discussion}

We have presented a tunable platform for investigating phase-non-reciprocal hopping dynamics between two spatially separated microwave oscillators. We explored uncharted parameter regimes where coupling significantly exceeds inherent losses at room temperature, by utilizing low-loss passive components and high-gain unidirectional amplifiers. We investigate the transmission behavior, LC, and synchronization phenomena that emerge when non-reciprocal amplifiers provide enough gain to compensate for loss in the hopping paths and to exceed the total loss in the system. The remarkable quantitative agreement between numerical, analytical, and experimental results demonstrates the effectiveness of our model in adequately describing the dynamics of the system.

Our platform holds significant potential across various fields in science and engineering, by facilitating tunable non-Hermitian and nonlinear dynamics. The ability to finely control the self-sustained LC frequency with phase opens new possibilities for cavity-magnonic and optomechanical systems \cite{Rodrigues_2021, rao_meterscale_2023, Qian2023, Barry2023}, as well as low-cost signal generators. Future characterization of the phase noise, output power, and linewidth, alongside comparisons with established microwave sources, could clarify whether the intrinsic phase-locking dynamics in the unstable regime contribute to phase noise suppression, enabling compact, frequency-tunable emitters with potentially enhanced spectral stability. Additionally, the observed frequency entrainment offers valuable insights for sensing applications \cite{xu_programmable_2020}, since the LC frequency shows enhanced sensitivity to the drive frequency at the onset of synchronization. Also promising for sensing is the non-Hermitian nature of our dimer: while, as noted, the undriven dynamics is normal, the combination of external drive and nonlinearity renders the linearized dynamics around the displaced steady state non-normal, enabling the exploration of exceptional points -- including in a nonlinear regime, where the sensitivity may potentially be enhanced
without sacrificing the signal-to-noise ratio as is typical of linear exceptional-point-based sensors \cite{Langbein2018, Loughlin2024, bai_observation_2024, carneyEP2025}. In turn, the sigmoid-like transmission profile we observe (Fig.\,\ref{fig:fig_2}\textbf{g}) offers opportunities for generating nonlinearities in analog neural networks across radio-to-optical frequencies \cite{liao_integrated_2023, Wanjura2024, Xia2024}. 

Furthermore, as mentioned, the dimer building block presented here can serve as a tunable edge in synthetic photonic materials, which can be extended to higher degrees of connectivity and enable non-planar geometries \cite{underwood_low-disorder_2012, owens_quarter-flux_2018, kollar_hyperbolic_2019}. Although the current system operates in a classical regime, its components can be adapted to cryogenic environments using parametric amplifiers \cite{castellanos-beltran_widely_2007, Abdo2013, macklin_nearquantum-limited_2015, aumentado_superconducting_2020} and YIG-based phase shifters, with promising implications for quantum information processing \cite{campagne-ibarcq_quantum_2020, Axline2018, Burkhart2021}. For example, one may envision that the coupling method introduced here may be useful in the context of novel implementations of driven-dissipative cat qubits \cite{grimm_stabilization_2020, Reglade2024}. 

Although the coupling that we introduced here is realized through a network of passive and active circuit elements, the dynamics are well captured by a two-mode coupled-mode model with phase-tunable non-reciprocal coupling. This phenomenological approach avoids unnecessary complexity while preserving generality and predictive power, and it reproduces the experimental behavior with excellent accuracy. Further characterization of cross-correlations between scattering elements across multiple phases and drive configurations, including simultaneous cavity driving, also represents a promising direction to deepen our understanding of the phase-nonreciprocal interaction between the coupled oscillators. Theoretically, we expect that a Lindblad master equation encompassing a novel combination of phase-dependent correlated loss and gain mechanisms will be needed to microscopically describe the system in a fully quantum Markovian regime. We leave this promising step to future investigation.


\section*{\label{sec:methods} 
Methods}

\subsection*{\label{sec:experimental-schematic} 
Full device description}

Our setup includes a tracking generator (SignalHound TG124A), synchronized with a spectrum analyzer (SignalHound SA124B), which serves as a scalar network analyzer and is utilized for collecting the data displayed in Fig.\,\ref{fig:fig_2}. Our dimer comprises two microwave cavities with mechanically tunable frequency and coupling rates. Each cavity is directly connected to an amplifier (Minicircuits CMA-83LN+), followed by a digital attenuator (Vaunix LDA-5018V) for tunable gain control. A phase shifter (Vaunix LPS-802) modifies the phase for the hopping path from cavity 2 to cavity 1. SMA coaxial pins (Minicircuits SM-SM50+) bridge connections between components where necessary. A 50 $\Omega$ terminator is connected to the driving port for LC experiments without an external drive. When a drive of precise strength and frequency is required, as shown in Fig.\,\ref{fig:fig_4}, an external signal generator (SignalCore SC5511A) is utilized.

\subsection*{\label{sec:hashmap} 
Calibration of $\Delta G$ and $\phi$}

The tunable parameters, namely, symmetric net gain in hopping, $\Delta G$, and relative phase, $\phi$, we use in the theoretical model are not quantities directly adjustable experimentally. Instead, they are inferred from actual experimental parameters, namely, the settings of the two digital attenuators $\Gamma_{1\to 2}$ and $\Gamma_{2\to 1}$ expressed in dB and the setting of the phase shifter $\phi_\text{exp}$ expressed in degrees. The calibration, or mapping, from $(\Gamma_{1\to 2}$, $\Gamma_{2\to 1}$, $\phi_\text{exp}) \mapsto (\Delta G,\phi)$ is achieved using a lookup table, referred to as the hash map, which was constructed from a series of calibrations. This hash map establishes the relationship between the experimental and model parameters, ensuring consistent $\Delta G$ for arbitrary $\phi$ (see Supplementary Information, Sec.\,VII).

\subsection*{\label{sec:exp-transmission-experiments} 
Weak-drive transmission experiments}

The experimental results shown in Fig. \ref{fig:fig_2}\textbf{b-g} were obtained using the following procedure. First, we calibrate our operational $\phi = 0$ by identifying the actual phase set on the device, $\phi_{\text{exp}}$, that makes the two peaks in Fig.\,\ref{fig:fig_2}\textbf{b} symmetric at the largest $\Delta G = 8.4$ dB value. Because the attenuator introduces phase shifts at different values of attenuation, $\Gamma$, we then iteratively adjust $\phi_{\text{exp}}$ to preserve this symmetry for all $\Delta G$ values displayed in Figs.\,\ref{fig:fig_2}\textbf{b,f}. At $\phi = \pi$, the primary goal was to update $\phi_{\text{exp}}$ to align the frequency of $S_{21}^{\text{max}}$ at $\omega_c/2\pi$ for different $\Gamma$. We then iteratively adjusted the digital attenuators to ensure a consistent $\Delta G$ value for all the data displayed in Fig.\,\ref{fig:fig_2}\textbf{d,g}. These additional phase offsets are mainly influenced by the inherent phase shift from increasing attenuation in the digital devices and by phase shifts due to amplifier saturation. These effects are challenging to characterize and control, thus requiring additional manual calibration. The spurious feature near $\omega_d / 2\pi \simeq$ 6.01 GHz in Fig. \ref{fig:fig_2}\textbf{d} arises from limit-cycle emission leaking into the scalar network analyzer during homodyne detection, not from a separate cavity mode, as confirmed by simulations in Fig. \ref{fig:fig_2}\textbf{e}. To extract the peak transmission ($S_{21}^\text{max}$) and linewidth (FWHM) in Fig.\ref{fig:fig_2}\textbf{f–g}, we fit each spectrum using double Lorentzians for $\phi = 0$ and a single Lorentzian for $\phi = \pi$. Above threshold, increased uncertainties in $S_{21}^{\text{max}}$ and FWHM arise due to spectral narrowing and the development of Fano-like asymmetries at $\phi = \pi$. The log-scale plot in Fig. \ref{fig:fig_2}\textbf{g} highlights these effects more clearly. Minor deviations in $S_{21}^\text{max}$ at low $\Delta G$ are also visible, possibly due to slightly lower attenuation than expected.

\subsection*{\label{sec:exp-limit-cycle} 
Experimental characterization of the limit cycle}

We used our hash-map for the LC characterization to ensure accurate $\Delta G$ across all $\phi$ values (Supplementary Information). The output port of cavity 2 is connected to a spectrum analyzer, and the emission spectrum is recorded for $\Delta G$ values ranging from 4.0 to 8.4 dB and $\phi$ from 0 to $2\pi$, with the previously determined $(\phi_{\text{exp}} \to \phi = 0)$ as the reference phase. A 50 $\Omega$ terminator is connected to the input port of cavity 1 to maintain the loading of the cavity during data collection. For each recorded emission spectrum, we extracted the highest amplitude and its corresponding frequency, identifying the LC amplitude and frequency for each $\Delta G$ and $\phi$, as shown in Fig.\,\ref{fig:fig_3}\textbf{a} and 3\textbf{d}, respectively (for more details see Supplementary Information Sec. IIE).

\subsection*{\label{sec:theoretical-framework} 
Sketch of the analytical analysis for the \\ undriven limit cycle}

When the system is not driven, $\epsilon = 0$, the relevant asymptotic behavior of the system can be solved for exactly.
In Sec.\,II of the Supplementary Information, we achieve this by introducing
the normal modes amplitudes $\beta_{\pm} \equiv \left(\pm e^{i \phi / 2} \alpha_{1} - \alpha_{2}\right) / \sqrt{2}$,
and working towards a suitable normal form for the EOMs of $\beta_{+}$, in polar coordinates $\beta_{+} = i R_{+} e^{i \theta_{+}}$, while assuming $\beta_{-} \equiv 0, \forall t$.

The radial EOM for $\dot{R}_{+}$, which is decoupled from that of $\dot{\theta}_{+}$, allows for a unique stable equilibrium point $R_{+} = R_{\textrm{LC}}$, when $R_{+} = 0$ becomes unstable. We describe this supercritical Hopf bifurcation explicitly,
and obtain the exact expression for $n_{\textrm{LC}} = R_{\textrm{LC}}^{2}$ in Eq.\,\eqref{eq:amp_lc}.
The uniqueness of $R_{\textrm{LC}}$ is a straightforward consequence of the monotonicity of $R_{+} \mapsto J(R_{+}^{2} / 2; \Delta G)$, with $J$ as defined in Eq.\,\eqref{eq:j_hop},
allowing applying its inverse on $( {|\alpha_{\textrm{sat}}|}^2, \infty ) \ni R_{\textrm{LC}}$.
Since ${|\alpha_{\textrm{sat}}|}^2 > 0$, and $J(R_{+}^{2} / 2; \Delta G)$ is perfectly constant for
$R_{+}^{2} < {|\alpha_{\textrm{sat}}|}^2$, the bifurcation described is technically no longer local,
since for the LC amplitude $R_{\textrm{LC}}$, we have that $R_{\textrm{LC}} > {|\alpha_{\textrm{sat}}|}^2$,
\emph{immediately} after $R_{+} = 0$ becomes unstable.

Next, we additionally derive an explicit expression for the LC linewidth by linearizing the radial EOM around $R_{+} = R_{\textrm{LC}}$. Lastly, the expression for the LC frequency in Eq.\,\eqref{eq:LC_frequency} is obtained by substituting $R_{+} = R_{\textrm{LC}}$ in the angular EOM for $\dot{\theta}_{+}$. This yields insight into the physical reason for $\delta \omega_{\textrm{LC}}$ not depending on $\Delta G$.

\subsection*{\label{sec:exp-synchronization} 
Synchronization experiments}

To investigate the synchronization dynamics between a self-sustained mode and an external drive, we connect the input port of the dimer to an external local oscillator (SignalCore SC5511A), which allows us to generate signals of precise frequency and power. Initially, we set the external drive to $\omega_d = \omega_c$ and record the emission spectra for different $\Delta G$ and $\phi$, adjusting the drive power, $P_d$, from 0 to 16 dBm. In each case, we identify regions where the spectrum displays two or more distinct peaks, indicating the coexistence of the drive and LC tones. These regions define the contours in Fig. \ref{fig:fig_4}\textbf{a}. We acknowledge the asymmetry observed in the contours of Fig. \ref{fig:fig_4}\textbf{a} and attribute it to minor experimental imperfections, such as slight mismatches in resonance conditions, parasitic phase shifts under amplifier saturation, or minor hysteresis in the phase tuning, which may cause the limit cycle to latch earlier on one side of $\phi = \pi$ (Supplementary Information Sec. III).

Next, we focus on characterizing the gap widening around $\phi = \pi$. We set the dimer parameters to $\Delta G = 8.4$ dB and $\phi = \pi$, which effectively fixes the LC frequency to $\omega_c/2\pi$. We then sweep $\omega_d/2\pi$ over an 8 MHz span around $\omega_c/2\pi$ at constant drive strengths of 0 and 4 dBm. The resulting spectra are displayed in Figs. \ref{fig:fig_4}\textbf{c-d}, respectively.

\subsection*{\label{sec:numerical-framework} 
Time-domain numerical analysis}

All simulated data in Figs.\,\ref{fig:fig_2}-\ref{fig:fig_4} are obtained using the Dormand-Prince method \cite{RK45} to solve Eq.\,\eqref{eq:dynamics_full}. To systematically analyze the time-domain traces, we applied the fast Fourier transform expressed as $y[k] \equiv \sum_{n=0}^{N-1} x[n] e^{-2\pi i {k n}/{N}}$, where $N = 1 \times 10^5$ is the total number of time samples. This transformation allows us to break down our time-domain signal into its frequency components. Hence, to numerically calculate $S_{21}$, we focus specifically on the DC component of the resulting spectra when the system is driven at $\omega_d/2\pi$. This DC component represents the average signal value over a given time interval and accurately represents the experimental homodyne detection, which measures the output signal at the drive frequency $\omega_d/2\pi$. 

Extending this analysis, we can also determine the frequency of the LC, $\delta \omega_{\text{LC}}^{\text{num}}/2\pi$, by extracting the dominant frequency in the computed spectra, $f(k_{\text{max}})$, namely, $\delta\omega_{\text{LC}}/2\pi = 2\pi f(k_{\text{max}})$, where $k_{\text{max}} = \arg\max_k |y[k]|$, and $f(k_{\text{max}}) = \tfrac{k_{\text{max}}}{N \Delta t}$. Here, $f(k_{\text{max}})$ defines the dominant frequency in the transformed spectra corresponding to the index $k_{\text{max}}$, and $\Delta t$ denotes the sampling interval. Numerical results in Figs.\,\ref{fig:fig_2}-\ref{fig:fig_4} were achieved by time-evolving the system in the range $0 \leq t \leq 10000 / \kappa_c$, under initial conditions set as $\alpha(0) = [10^7, 0, 10^7, 0]$. We excluded the initial $20\%$ of the data to ignore transient effects, and whenever $|\alpha_2|^2 < 0.001\% |\alpha_{\text{sat}}|^2$, to avoid spurious sampling effects from numerical traces corresponding to the vacuum solution. We convert the photon numbers to absolute power using $P\ [\text{W}] = \hbar \omega_c |\alpha_2|^2 \kappa_{\text{out}}$, and then calculate the amplitude in dBm using the conversion $10 \log_{10} \left(P / 1\ \text{mW}\right)$. We also refer to the Supplementary Information Sec.\,ID for more details regarding the numerical parameters used in simulations.

\section*{\label{sec:data-availability} Data Availability}

All data that support the findings of this study are publicly available in the GitHub repository listed below and archived in Zenodo to ensure reproducibility \cite{salcedogallo_repo}. 

\section*{\label{sec:code-availability} Code Availability}

The code that supports the findings of this study are available at \url{https://www.github.com/jussalcedoga-dartmouth/nh_nl_dynamics_dimer} \cite{salcedogallo_repo}.



\begin{thebibliography}{76}

\bibitem{Ashida}
Zongping~Gong Yuto~Ashida and Masahito Ueda.
\newblock Non-hermitian physics.
\newblock {\em Adv. Phys.}, 69(3):249--435, 2020.

\bibitem{Feshbach}
H.~Feshbach, C.~E. Porter, and V.~F. Weisskopf.
\newblock Model for nuclear reactions with neutrons.
\newblock {\em Phys. Rev.}, 96:448, 1954.

\bibitem{HN}
Naomichi Hatano and David~R. Nelson.
\newblock Localization transitions in non-{H}ermitian quantum mechanics.
\newblock {\em Phys. Rev. Lett.}, 77:570--573, 1996.

\bibitem{El-Ganainy07}
R.~El-Ganainy, K.~G. Makris, D.~N. Christodoulides, and Ziad~H. Musslimani.
\newblock Theory of coupled optical {PT}-symmetric structures.
\newblock {\em Opt. Lett.}, 32(17):2632--2634, 2007.

\bibitem{Graefe_2010}
Eva~Maria Graefe, Michael Honing, and Hans~Jurgen Korsch.
\newblock Classical limit of non-hermitian quantum dynamics as generalized canonical structure.
\newblock {\em J. Phys. A: Math. Theor.}, 43(7):075306, 2010.

\bibitem{el-ganainy_non-hermitian_2018}
Ramy El-Ganainy, Konstantinos~G. Makris, Mercedeh Khajavikhan, Ziad~H. Musslimani, Stefan Rotter, and Demetrios~N. Christodoulides.
\newblock Non-{H}ermitian physics and {PT} symmetry.
\newblock {\em Nat. Phys.}, 14(1):11--19, 2018.

\bibitem{Bender_1998}
Carl~M. Bender and Stefan Boettcher.
\newblock Real spectra in non-{H}ermitian hamiltonians having $\mathcal{P}\mathcal{T}$ symmetry.
\newblock {\em Phys. Rev. Lett.}, 80:5243--5246, 1998.

\bibitem{Bender_2007}
Carl~M Bender.
\newblock Making sense of non-{H}ermitian {H}amiltonians.
\newblock {\em Rep. Prog. Phys.}, 70(6):947, 2007.

\bibitem{Salamo}
A.~Guo, G.~J. Salamo, D.~Duchesne, R.~Morandotti, M.~Volatier-Ravat, V.~Aimez, G.~A. Siviloglou, and D.~N. Christodoulides.
\newblock Observation of $\mathcal{P}\mathcal{T}$-symmetry breaking in complex optical potentials.
\newblock {\em Phys. Rev. Lett.}, 103:093902, 2009.

\bibitem{bittner_p_2012}
S.~Bittner, B.~Dietz, U.~G\"{u}nther, H.~L. Harney, M.~Miski-Oglu, A.~Richter, and F.~Sch\"{a}fer.
\newblock {P}{T} {S}ymmetry and {S}pontaneous {S}ymmetry {B}reaking in a {M}icrowave {B}illiard.
\newblock {\em Phys. Rev. Lett.}, 108:024101, 2012.

\bibitem{Feng2017}
Liang Feng, Ramy El-Ganainy, and Li~Ge.
\newblock Non-hermitian photonics based on parity--time symmetry.
\newblock {\em Nat. Photonics}, 11(12):752--762, 2017.

\bibitem{Jin2024}
Chanju Kim, Xinda Lu, Deming Kong, Nuo Chen, Yuntian Chen, Leif~Katsuo Oxenl{\o}we, Kresten Yvind, Xinliang Zhang, Lan Yang, Minhao Pu, and Jing Xu.
\newblock Parity-time symmetry enabled ultra-efficient nonlinear optical signal processing.
\newblock {\em eLight}, 4(1):6, 2024.

\bibitem{Breuer2009}
H.~P. Breuer and F.~Petruccione.
\newblock {\em The {T}heory of {O}pen {Q}uantum {S}ystems}.
\newblock Oxford University Press, Great Clarendon Street, 2002.

\bibitem{Lindblad1976}
G.~Lindblad.
\newblock On the generators of quantum dynamical semigroups.
\newblock {\em Commun. Math. Phys.}, 48:119--130, 1976.

\bibitem{blazoit86}
J.~P. Blaizot and G.~Ripka.
\newblock {\em Quantum Theory of Finite Systems}.
\newblock The MIT Press, Cambridge, Massachusetts, 1986.

\bibitem{ClerkBKC}
A.~McDonald, T.~Pereg-Barnea, and A.~A. Clerk.
\newblock Phase-dependent chiral transport and effective non-{H}ermitian dynamics in a bosonic kitaev-majorana chain.
\newblock {\em Phys. Rev. X}, 8:041031, 2018.

\bibitem{Decon}
V.~P. Flynn, E.~Cobanera, and L.~Viola.
\newblock Deconstructing effective non-{H}ermitian dynamics in quadratic bosonic hamiltonians.
\newblock {\em New J. Phys.}, 22:083004, 2020.

\bibitem{Mariam}
Mariam Ughrelidze, Vincent~P. Flynn, Emilio Cobanera, and Lorenza Viola.
\newblock Interplay of finite- and infinite-size stability in quadratic bosonic lindbladians.
\newblock {\em Phys. Rev. A}, 110:032207, 2024.

\bibitem{metelmann_nonreciprocal_2015}
A.~Metelmann and A.~A. Clerk.
\newblock Nonreciprocal photon transmission and amplification via reservoir engineering.
\newblock {\em Phys. Rev. X}, 5(2):021025, 2015.

\bibitem{Painter}
Kejie Fang, Jie Luo, Anja Metelmann, Matthew~H. Matheny, Florian Marquardt, Aashish~A. Clerk, and Oskar Painter.
\newblock Generalized non-reciprocity in an optomechanical circuit via synthetic magnetism and reservoir engineering.
\newblock {\em Nat. Phys.}, 13(5):465--471, 2017.

\bibitem{PorrasTopoAmp}
D.~Porras and S.~Fernandez-Lorenzo.
\newblock Topological amplification in photonic lattices.
\newblock {\em Phys. Rev. Lett.}, 122:143901, 2019.

\bibitem{Wanjura2020}
C.~Wanjura, M.~Brunelli, and A.~Nunnenkamp.
\newblock Topological framework for directional amplification in driven-dissipative cavity arrays.
\newblock {\em Nat. Commun.}, 11(1):3149, 2020.

\bibitem{mcdonald_exponentially-enhanced_2020}
Alexander McDonald and Aashish~A. Clerk.
\newblock Exponentially-enhanced quantum sensing with non-{H}ermitian lattice dynamics.
\newblock {\em Nat. Commun.}, 11(1):5382, 2020.

\bibitem{Wiersig}
Jan Wiersig.
\newblock Prospects and fundamental limits in exceptional point-based sensing.
\newblock {\em Nat. Commun.}, 11(1):2454, 2020.

\bibitem{CarusottoRMP}
Tomoki Ozawa, Hannah~M. Price, Alberto Amo, Nathan Goldman, Mohammad Hafezi, Ling Lu, Mikael~C. Rechtsman, David Schuster, Jonathan Simon, Oded Zilberberg, and Iacopo Carusotto.
\newblock Topological photonics.
\newblock {\em Rev. Mod. Phys.}, 91:015006, 2019.

\bibitem{Flebus}
Hilary~M. Hurst and Benedetta Flebus.
\newblock Non-{H}ermitian physics in magnetic systems.
\newblock {\em J. Appl. Phys.}, 132(22):220902, 2022.

\bibitem{Vitelli}
Michel Fruchart, Ryo Hanai, Peter~B. Littlewood, and Vincenzo Vitelli.
\newblock Non-reciprocal phase transitions.
\newblock {\em Nature}, 592(7854):363--369, 2021.

\bibitem{flynn_topology_2021}
Vincent~P. Flynn, Emilio Cobanera, and Lorenza Viola.
\newblock Topology by dissipation: Majorana bosons in metastable quadratic markovian dynamics.
\newblock {\em Phys. Rev. Lett.}, 127:245701, 2021.

\bibitem{flynn_2023}
Vincent~P. Flynn, Emilio Cobanera, and Lorenza Viola.
\newblock Topological zero modes and edge symmetries of metastable markovian bosonic systems.
\newblock {\em Phys. Rev. B}, 108:214312, 2023.

\bibitem{OkumaSato}
Nobuyuki Okuma and Masatoshi Sato.
\newblock Non-{H}ermitian topological phenomena: {A} review.
\newblock {\em Annu. Rev. Condens. Matter Phys.}, 14:83--107, 2023.

\bibitem{Liu2023}
Tongjun Liu, Jun~Yu Ou, Kevin~F. MacDonald, and Nikolay~I. Zheludev.
\newblock Photonic metamaterial analogue of a continuous time crystal.
\newblock {\em Nat. Phys.}, 19(7):986--991, 2023.

\bibitem{Raskatla2024}
V.~Raskatla, T.~Liu, J.~Li, K.~F. MacDonald, and Nikolay~I. Zheludev.
\newblock Continuous space-time crystal state driven by nonreciprocal optical forces.
\newblock {\em Phys. Rev. Lett.}, 133:136202, 2024.

\bibitem{brzobohaty_synchronization_2023}
O.~Brzobohat{\'y}, M.~Duchan, P.~Jakl, J.~Jezek, M.~Siler, P.~Zemanek, and S.~H. Simpson.
\newblock Synchronization of spin-driven limit cycle oscillators optically levitated in vacuum.
\newblock {\em Nat. Commun.}, 14(1):5441, 2023.

\bibitem{reisenbauer_non-Hermitian_2024}
Manuel Reisenbauer, Henning Rudolph, Livia Egyed, Klaus Hornberger, Anton~V. Zasedatelev, Murad Abuzarli, Benjamin~A. Stickler, and Uro{\v{s}} Deli{\'{c}}.
\newblock Non-hermitian dynamics and non-reciprocity of optically coupled nanoparticles.
\newblock {\em Nat. Phys.}, 20(10):1629--1635, 2024.

\bibitem{liska_pt-like_2024}
Vojt{\v{e}}ch Li{\v{s}}ka, Tereza Zem{\'a}nkov{\'a}, Petr J{\'a}kl, Martin {\v{S}}iler, Stephen~H. Simpson, Pavel Zem{\'a}nek, and Oto Brzobohat{\'y}.
\newblock Pt-like phase transition and limit cycle oscillations in non-reciprocally coupled optomechanical oscillators levitated in vacuum.
\newblock {\em Nat. Phys.}, 20(10):1622--1628, 2024.

\bibitem{BKCExp}
Jesse~J. Slim, Clara~C. Wanjura, Matteo Brunelli, Javier del Pino, Andreas Nunnenkamp, and Ewold Verhagen.
\newblock Optomechanical realization of the bosonic {K}itaev chain.
\newblock {\em Nature}, 627(8005):767--771, 2024.

\bibitem{Pedergnana_2024}
Tiemo Pedergnana, Abel Faure-Beaulieu, Romain Fleury, and Nicolas Noiray.
\newblock Loss-compensated non-reciprocal scattering based on synchronization.
\newblock {\em Nat. Commun.}, 15(1):7436, 2024.

\bibitem{rao_meterscale_2023}
Jinwei Rao, C.~Y. Wang, Bimu Yao, Z.~J. Chen, K.~X. Zhao, and Wei Lu.
\newblock Meterscale {Strong} {Coupling} between {Magnons} and {Photons}.
\newblock {\em Phys. Rev. Lett.}, 131:106702, 2023.

\bibitem{owens_quarter-flux_2018}
Clai Owens, Aman LaChapelle, Brendan Saxberg, Brandon~M. Anderson, Ruichao Ma, Jonathan Simon, and David~I. Schuster.
\newblock Quarter-flux {H}ofstadter lattice in a qubit-compatible microwave cavity array.
\newblock {\em Phys. Rev. A}, 97(1):013818, 2018.

\bibitem{owens_chiral_2022}
John~Clai Owens, Margaret~G. Panetta, Brendan Saxberg, Gabrielle Roberts, Srivatsan Chakram, Ruichao Ma, Andrei Vrajitoarea, Jonathan Simon, and David~I. Schuster.
\newblock Chiral cavity quantum electrodynamics.
\newblock {\em Nat. Phys.}, 18(9):1048--1052, 2022.

\bibitem{Nunnenkamp_2011}
A~Nunnenkamp, Jens Koch, and S~M Girvin.
\newblock Synthetic gauge fields and homodyne transmission in {J}aynes-{C}ummings lattices.
\newblock {\em New J. Phys.}, 13(9):095008, 2011.

\bibitem{Houck2012}
Andrew~A. Houck, Hakan~E. Tureci, and Jens Koch.
\newblock On-chip quantum simulation with superconducting circuits.
\newblock {\em Nat. Phys.}, 8(4):292--299, 2012.

\bibitem{kollar_hyperbolic_2019}
Alicia~J. Kollar, Mattias Fitzpatrick, and Andrew~A. Houck.
\newblock Hyperbolic lattices in circuit quantum electrodynamics.
\newblock {\em Nature}, 571(7763):45--50, 2019.

\bibitem{Carusotto2020}
Iacopo Carusotto, Andrew~A. Houck, Alicia~J. Kollar, Pedram Roushan, David~I. Schuster, and Jonathan Simon.
\newblock Photonic materials in circuit quantum electrodynamics.
\newblock {\em Nat. Phys.}, 16(3):268--279, 2020.

\bibitem{Holmes}
J.~Guckenheimer and P.~Holmes.
\newblock {\em Nonlinear {O}scillations, {D}ynamical {S}ystems, and {B}ifurcations of {V}ector {F}ields}.
\newblock Springer Verlag, 1983.

\bibitem{Hayashi1964}
Chihiro Hayashi.
\newblock {\em Nonlinear {O}scillations in {P}hysical {S}ystems}.
\newblock McGraw-Hill, revised and enlarged edition edition, 1964.

\bibitem{Balanov2009}
A.~Balanov, N.~Janson, D.~Postnov, and Olga Sosnovtseva.
\newblock {\em Synchronization: From Simple to Complex}.
\newblock Springer, 2009.

\bibitem{Sadeghpour2013}
Tony~E. Lee and H.~R. Sadeghpour.
\newblock Quantum synchronization of quantum van der {P}ol oscillators with trapped ions.
\newblock {\em Phys. Rev. Lett.}, 111:234101, 2013.

\bibitem{Walter2014}
Stefan Walter, Andreas Nunnenkamp, and Christoph Bruder.
\newblock Quantum synchronization of a driven self-sustained oscillator.
\newblock {\em Phys. Rev. Lett.}, 112:094102, 2014.

\bibitem{Walter2015}
Stefan Walter, Andreas Nunnenkamp, and Christoph Bruder.
\newblock Quantum synchronization of two van der pol oscillators.
\newblock {\em Ann. Phys.}, 527:131--138, 2015.

\bibitem{Zhang_2024}
Chunlei Zhang, Mun Kim, Jianbo Wang, and Can~Ming Hu.
\newblock Van der {P}ol--{D}uffing oscillator and its application to gain-driven light-matter interaction.
\newblock {\em Phys. Rev. Appl.}, 22:014034, 2024.

\bibitem{Solanski2025}
T.~Kehrer, C.~Bruder, and P.~Solanki.
\newblock Quantum synchronization of twin limit-cycle oscillators.
\newblock {\em arXiv:2502.21122}, 2025.

\bibitem{backscatter_2020}
Nils~T. Otterstrom, Shai Gertler, Yishu Zhou, Eric~A. Kittlaus, Ryan~O. Behunin, Michael Gehl, Andrew~L. Starbuck, Christina~M. Dallo, Andrew~T. Pomerene, Douglas~C. Trotter, Anthony~L. Lentine, and Peter~T. Rakich.
\newblock Backscatter-immune injection-locked {B}rillouin laser in silicon.
\newblock {\em Phys. Rev. Appl.}, 14:044042, 2020.

\bibitem{Rodrigues_2021}
Caique~C. Rodrigues, Cau{\^e}~M. Kersul, Andr{\'e}~G. Primo, Michal Lipson, Thiago P.~Mayer Alegre, and Gustavo~S. Wiederhecker.
\newblock Optomechanical synchronization across multi-octave frequency spans.
\newblock {\em Nat. Commun.}, 12(1):5625, 2021.

\bibitem{Qian2023}
Jie Qian, C.~H. Meng, J.~W. Rao, Z.~J. Rao, Zhenghua An, Yongsheng Gui, and C.-M. Hu.
\newblock Non-hermitian control between absorption and transparency in perfect zero-reflection magnonics.
\newblock {\em Nat. Commun.}, 14(1):3437, 2023.

\bibitem{Barry2023}
John~F. Barry, Reed~A. Irion, Matthew~H. Steinecker, Daniel~K. Freeman, Jessica~J. Kedziora, Reginald~G. Wilcox, and Danielle~A. Braje.
\newblock Ferrimagnetic oscillator magnetometer.
\newblock {\em Phys. Rev. Appl.}, 19:044044, 2023.

\bibitem{xu_programmable_2020}
Liu Xu, Shudong Wang, Zhuangde Jiang, and Xueyong Wei.
\newblock Programmable synchronization enhanced mems resonant accelerometer.
\newblock {\em Microsyst. Nanoeng.}, 6(1):63, 2020.

\bibitem{Langbein2018}
W.~Langbein.
\newblock No exceptional precision of exceptional-point sensors.
\newblock {\em Phys. Rev. A}, 98:023805, 2018.

\bibitem{Loughlin2024}
Hudson Loughlin and Vivishek Sudhir.
\newblock Exceptional-point sensors offer no fundamental signal-to-noise ratio enhancement.
\newblock {\em Phys. Rev. Lett.}, 132:243601, 2024.

\bibitem{bai_observation_2024}
Kai Bai, Tian~Rui Liu, Liang Fang, Jia~Zheng Li, Chen Lin, Duanduan Wan, and Meng Xiao.
\newblock Observation of {Nonlinear} {Exceptional} {Points} with a {Complete} {Basis} in {Dynamics}.
\newblock {\em Phys. Rev. Lett.}, 132(7):073802, 2024.

\bibitem{carneyEP2025}
Alexander~S. Carney, Juan~S. Salcedo-Gallo, Salil~K. Bedkihal, and Mattias Fitzpatrick.
\newblock Unification of exceptional points and transmission peak degeneracies in a highly tunable magnon-photon dimer, 2025.

\bibitem{liao_integrated_2023}
Kun Liao, Tianxiang Dai, Qiuchen Yan, Xiaoyong Hu, and Qihuang Gong.
\newblock Integrated {Photonic} {Neural} {Networks}: {O}pportunities and challenges.
\newblock {\em ACS Photonics}, 10:2001--2010, 2023.

\bibitem{Wanjura2024}
Clara~C. Wanjura and Florian Marquardt.
\newblock Fully nonlinear neuromorphic computing with linear wave scattering.
\newblock {\em Nat. Phys.}, 20(9):1434--1440, 2024.

\bibitem{Xia2024}
Fei Xia, Kyungduk Kim, Yaniv Eliezer, SeungYun Han, Liam Shaughnessy, Sylvain Gigan, and Hui Cao.
\newblock Nonlinear optical encoding enabled by recurrent linear scattering.
\newblock {\em Nat. Photonics}, 18(10):1067--1075, 2024.

\bibitem{underwood_low-disorder_2012}
D.~L. Underwood, W.~E. Shanks, Jens Koch, and A.~A. Houck.
\newblock Low-disorder microwave cavity lattices for quantum simulation with photons.
\newblock {\em Phys. Rev. A}, 86(2):023837, 2012.

\bibitem{castellanos-beltran_widely_2007}
M.~Castellanos-Beltran and K.~W. Lehnert.
\newblock Widely tunable parametric amplifier based on a superconducting quantum interference device array resonator.
\newblock {\em Appl. Phys. Lett.}, 91(8):083509, 2007.

\bibitem{Abdo2013}
Baleegh Abdo, Katrina Sliwa, Luigi Frunzio, and Michel Devoret.
\newblock Directional amplification with a {J}osephson circuit.
\newblock {\em Phys. Rev. X}, 3:031001, 2013.

\bibitem{macklin_nearquantum-limited_2015}
C.~Macklin, K.~O'Brien, D.~Hover, M.~E. Schwartz, V.~Bolkhovsky, X.~Zhang, W.~D. Oliver, and I.~Siddiqi.
\newblock A near-quantum-limited {J}osephson traveling-wave parametric amplifier.
\newblock {\em Science}, 350(6258):307--310, 2015.

\bibitem{aumentado_superconducting_2020}
Jose Aumentado.
\newblock Superconducting {Parametric} {Amplifiers}: {The} {State} of the {Art} in {Josephson} {Parametric} {Amplifiers}.
\newblock {\em IEEE Microw. Mag.}, 21(8):45--59, 2020.

\bibitem{campagne-ibarcq_quantum_2020}
P.~Campagne-Ibarcq, A.~Eickbusch, S.~Touzard, E.~Zalys-Geller, N.~E. Frattini, V.~V. Sivak, P.~Reinhold, S.~Puri, S.~Shankar, R.~J. Schoelkopf, L.~Frunzio, M.~Mirrahimi, and M.~H. Devoret.
\newblock Quantum error correction of a qubit encoded in grid states of an oscillator.
\newblock {\em Nature}, 584(7821):368--372, 2020.

\bibitem{Axline2018}
Christopher~J. Axline, Luke~D. Burkhart, Wolfgang Pfaff, Mengzhen Zhang, Kevin Chou, Philippe Campagne-Ibarcq, Philip Reinhold, Luigi Frunzio, S.~M. Girvin, Liang Jiang, M.~H. Devoret, and R.~J. Schoelkopf.
\newblock On-demand quantum state transfer and entanglement between remote microwave cavity memories.
\newblock {\em Nat. Phys.}, 14(7):705--710, 2018.

\bibitem{Burkhart2021}
Luke~D. Burkhart, James~D. Teoh, Yaxing Zhang, Christopher~J. Axline, Luigi Frunzio, M.H. Devoret, Liang Jiang, S.M. Girvin, and R.J. Schoelkopf.
\newblock Error-detected state transfer and entanglement in a superconducting quantum network.
\newblock {\em PRX Quantum}, 2:030321, 2021.

\bibitem{grimm_stabilization_2020}
A.~Grimm, N.~E. Frattini, S.~Puri, S.~O. Mundhada, S.~Touzard, M.~Mirrahimi, S.~M. Girvin, S.~Shankar, and M.~H. Devoret.
\newblock Stabilization and operation of a {Kerr}-cat qubit.
\newblock {\em Nature}, 584(7820):205--209, 2020.

\bibitem{Reglade2024}
U.~R{\'e}glade, A.~Bocquet, R.~Gautier, J.~Cohen, A.~Marquet, E.~Albertinale, N.~Pankratova, M.~Hall{\'e}n, F.~Rautschke, L.~A. Sellem, P.~Rouchon, A.~Sarlette, M.~Mirrahimi, P.~Campagne-Ibarcq, R.~Lescanne, S.~Jezouin, and Z.~Leghtas.
\newblock Quantum control of a cat qubit with bit-flip times exceeding ten seconds.
\newblock {\em Nature}, 629(8013):778--783, 2024.

\bibitem{RK45}
J.~R. Dormand and P.~J. Prince.
\newblock A family of embedded {R}unge-{K}utta formulae.
\newblock {\em J. Comput. Appl. Math.}, 6:19, 1980.

\bibitem{salcedogallo_repo}
J.~S. Salcedo-Gallo, M.~Burgelman, V.~P. Flynn, A.~S. Carney, M.~Hamdan, T.~Gerg, D.~C. Smallwood, L.~Viola, and M.~Fitzpatrick.
\newblock Demonstration of a tunable non-hermitian nonlinear microwave dimer.
\newblock \url{https://doi.org/10.5281/zenodo.15887711}, 2025.
\newblock GitHub repository: \texttt{nh\_nl\_dynamics\_dimer}.

\end{thebibliography}



\begin{acknowledgments}
We are grateful to Joe Poissant for his exceptional support in designing and fabricating the 3D microwave cavities. It is a pleasure to thank Salil K. Bedkihal for a critical reading of and valuable input to this manuscript, and Yikang Zhang for useful discussions. We are especially indebted to referees for providing insightful comments that substantially improved our analysis and interpretation of the phase-locking synchronization phenomena in our system. Startup funds from the Thayer School of Engineering, Dartmouth College, supported this work. We gratefully acknowledge support from DARPA Young Faculty Award No.\,D23AP00192 (to MF) and from the NSF through Grants No.\,PHY-2013974, PHY-2412555 (to LV) as well as DGE-2125733 (to ASC) through the Traineeship in Transformative Research and Graduate Education in Sensor Science, Technology, and Innovation.
\end{acknowledgments}

\section*{\label{sec:author-contribution} 
Author Contributions}

J.S.S.G., A.S.C., M.H., D.C.S., and M.F. conceived and designed the experiments, while J.S.S.G. performed the final set of experiments and calibrations. T.G. and D.C.S. conducted HFSS simulations on the initial designs for the individual cavities. The theoretical model was developed by M.B., V.P.F., A.S.C., J.S.S.G., L.V., and M.F., with M.B. and V.P.F. developing the analytical derivation of the undriven limit cycle and assisted in data analysis and interpretation, receiving feedback from L.V. and M.F. J.S.S.G. implemented and performed the numerical simulations. All authors jointly validated the results and participated in writing of the manuscript. L.V. and M.F. supervised the project.

\section*{\label{sec:competing-interests} 
Competing Interests}

The authors declare no competing interests.

\section*{\label{sec:additional-information} 
Additional information}

\subsection*{Supplementary Information} In this pre-print version, the Main Manuscript and Supplementary Information are provided as a single file for immediate access.

\end{document}


\title{Supplementary Information: \\
Demonstration of a Tunable Non-Hermitian Nonlinear Microwave Dimer}

\author{Juan S. Salcedo-Gallo}
\affiliation{\mbox{Thayer School of Engineering, Dartmouth College, 15 Thayer Drive, Hanover, New Hampshire 03755, USA}}

\author{Michiel Burgelman}
\affiliation{\mbox{Department of Physics and Astronomy, Dartmouth College, 6127 Wilder Laboratory, Hanover, New Hampshire 03755, USA}}

\author{Vincent P. Flynn}
\affiliation{\mbox{Department of Physics and Astronomy, Dartmouth College, 6127 Wilder Laboratory, Hanover, New Hampshire 03755, USA}}

\author{Alexander S. Carney}
\affiliation{\mbox{Thayer School of Engineering, Dartmouth College, 15 Thayer Drive, Hanover, New Hampshire 03755, USA}}

\author{\mbox{Majd Hamdan}}
\affiliation{\mbox{Thayer School of Engineering, Dartmouth College, 15 Thayer Drive, Hanover, New Hampshire 03755, USA}}

\author{Tunmay Gerg}
\affiliation{\mbox{Department of Physics and Astronomy, Dartmouth College, 6127 Wilder Laboratory, Hanover, New Hampshire 03755, USA}}

\author{Daniel C. Smallwood}
\affiliation{\mbox{Thayer School of Engineering, Dartmouth College, 15 Thayer Drive, Hanover, New Hampshire 03755, USA}}

\author{Lorenza Viola}
\affiliation{\mbox{Department of Physics and Astronomy, Dartmouth College, 6127 Wilder Laboratory, Hanover, New Hampshire 03755, USA}}

\author{Mattias Fitzpatrick}
\affiliation{\mbox{Thayer School of Engineering, Dartmouth College, 15 Thayer Drive, Hanover, New Hampshire 03755, USA}}
\affiliation{\mbox{Department of Physics and Astronomy, Dartmouth College, 6127 Wilder Laboratory, Hanover, New Hampshire 03755, USA}}

\maketitle


\tableofcontents

\section{Theoretical Model}

\subsection{Characterizing the dynamical matrix: On-site dissipation rates}

In this section, we motivate in more detail each term for the linear model 
presented in the main manuscript. As mentioned therein, we characterize the system using the quantity $\Delta G = G_0 - \Gamma$, which effectively represents the net gain in hopping. Specifically, 
tunable hopping coefficient at low power is obtained by $J_0(\Delta G) = 10^{\Delta G/20} \kappa_c$. To describe the dynamics under these conditions, we employ the following semi-classical EOMs:
\begin{eqnarray}
\label{eq:dynamics}
\dot{\bm{\alpha}} &=& A_{0}
\bm{\alpha} + \epsilon\textbf{B},  
\end{eqnarray}
expressed in terms of state variables $\bm{\alpha}\equiv \begin{bmatrix}
\alpha_1 \: \alpha_2
\end{bmatrix}^T$, $\textbf{B} 
\equiv \begin{bmatrix} 1 \:0
\end{bmatrix}^T$, respectively. We defined the linear drive strength as $\epsilon = \sqrt{\kappa_{\text{in}} P_{\text{in}} / \hbar \omega_d}$, where $P_{\text{in}}$ is the input drive power in watts, converted from dBm using $P_{\text{in}} = 10^{(P_d - 30)/10}$, and $\omega_d$ is the angular frequency of the external drive tone. $A_0$ is a dynamical matrix takes the form 
\begin{align}
\label{eq:general_form}
A_0 =  \begin{bmatrix}
-i(\omega_c - \omega_d) - \kappa & -iJ_0(\Delta G) e^{-i\phi} \\ -iJ_0(\Delta G) & -i(\omega_c -\omega_d) - \kappa
\end{bmatrix}.
\end{align} 
As mentioned in the main text, $\omega_c$ represents the common angular resonance frequency of the cavities, $\kappa$ denotes the total effective intra-cavity dissipation rate, which are taken to be the same for both cavities, simplifying the analytical derivations.
This is a good approximation, as can be seen from Supplementary Tables ~\ref{tab:paramtab},~\ref{tab:diss_rates_1} and~\ref{tab:diss_rates_2} (also see Sec.\,\ref{sec:experimental_parameters_cavity}).
$J_0(\Delta G)$ characterizes the effective coupling strength, with $\phi$ modulating the relative phase of the propagating signal in the backward direction. According to Eq.\,\eqref{eq:dynamics}, the system converges to a unique stable equilibrium when the maximum of the real part of the eigenvalues of 
$A_0(\Delta G, \phi)$, denoted as $\max \text{Re} [\sigma(A_0(\Delta G, \phi))]$, is strictly negative.
This condition serves as a precise criterion for stability of Eq.\,\eqref{eq:dynamics} (i.e. stability of the dynamical matrix $A_0$), 
given by the following equation:
\begin{align}
\label{eq:stability_linear}
\frac{J_0(\Delta G)}{\kappa} \sin(\phi/2) < 1.
\end{align}

Initially, we considered two candidate functions to represent the intra-cavity dissipation rate $\kappa$ in our system, namely, 
\begin{align}
\label{eq:base_kappa}
\kappa & \equiv \kappa_0 = \kappa_{\text{int}} + \kappa_{\text{in/out}} + \kappa_c, 
\\
\label{eq:diss_kappa_deltaG}
\kappa & \equiv \kappa_0(\Delta G) = 2(\kappa_{\text{int}} + \kappa_{\text{in/out}} + \kappa_c) - J_0(\Delta G), 
\end{align} 
where the term $\kappa_{\text{int}}$ represents the intrinsic dissipation of each oscillator, and $\kappa_{\text{in}}$, $\kappa_{\text{out}}$, and $\kappa_{c}$ account for the input, output, and coupling rates, respectively. Each additional port contributes to oscillator leakage and, therefore, increases on-site loss. Note that we assumed here that
$    \kappa_{\textrm{int}, 1} = \kappa_{\textrm{int}, 2},$ and 
$    \kappa_{\textrm{in}} = \kappa_{\textrm{out}}.$
We consider the average values between the corresponding measured parameters for each cavity: $\kappa_{\text{int}}/2\pi = 4.05$ MHz,  and $\kappa_{\text{in/out}}/2\pi = 2.4$ MHz.
\begin{figure}[!htb]
\centering
\includegraphics[width=.6\textwidth]{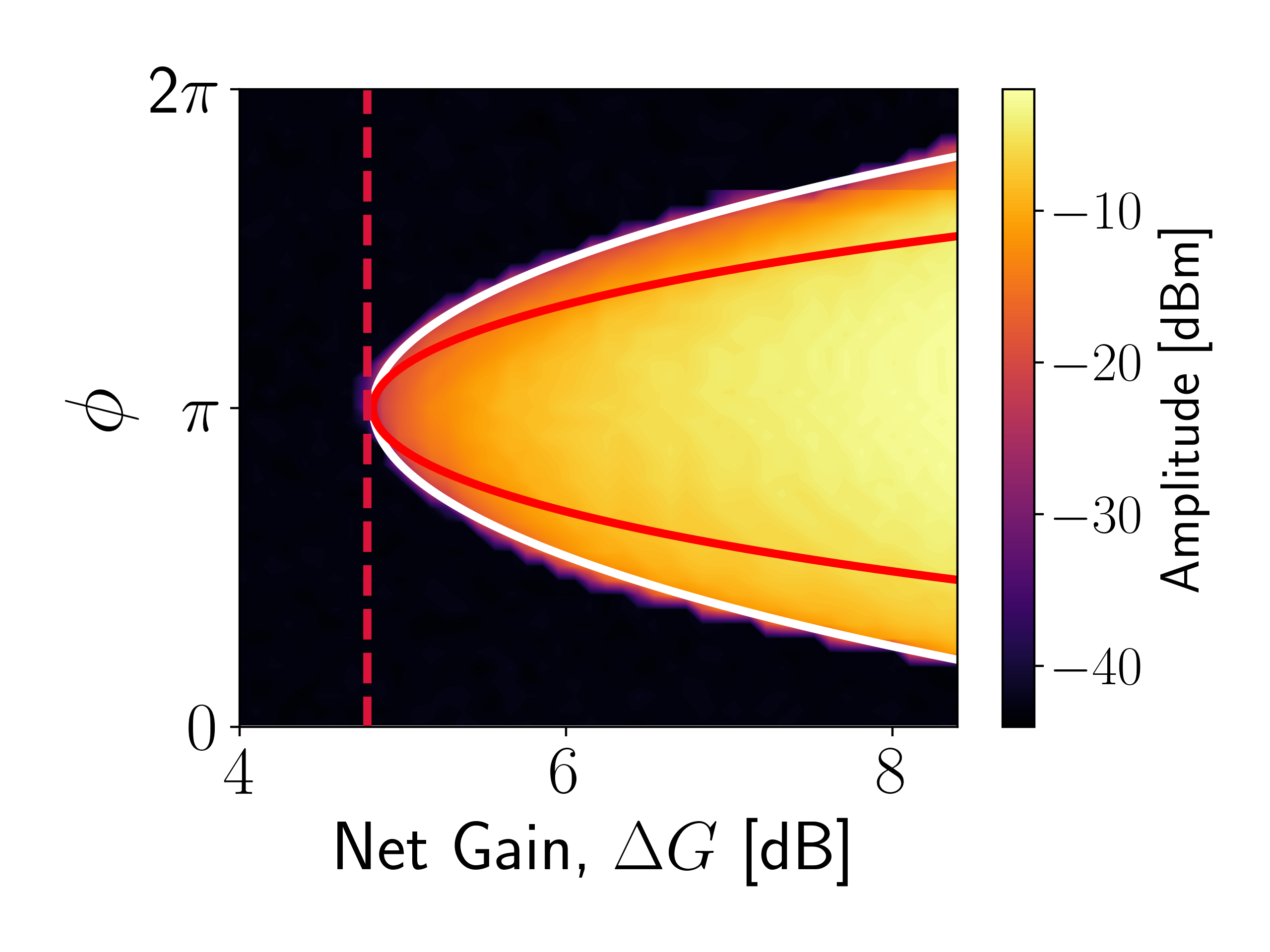}
\vspace*{-4mm}
\caption{\textbf{Stability phase diagram for the dynamical system}. Experimentally measured LC amplitude as a function of $\phi$ and $\Delta G$, with the theoretically predicted stability phase boundary for the dynamical matrix $A_0$ in Eq.\,\eqref{eq:general_form}, using two candidate on-site dissipation functions: 
$\kappa_0$ 
(red solid line)
[Eq.\,\eqref{eq:base_kappa}] 
and $\kappa_0(\Delta G)$ 
(white solid line)
[Eq.\,\eqref{eq:diss_kappa_deltaG}]. 
The vertical dashed line represents the $\Delta G$ value at which the system first becomes unstable at $\phi = \pi$.}
\label{fig:phase_diag}
\end{figure}

In the main manuscript, we showed that once the 
equilibrium point becomes unstable, the system converges toward a self-sustained limit cycle (LC) in phase space. \suppfigref{fig:phase_diag} illustrates the onset of instability as determined by Eq.\,\eqref{eq:stability_linear} for the two on-site dissipation models discussed above: the red solid line represents Eq.\,\eqref{eq:base_kappa}, while the white solid line represents Eq.\,\eqref{eq:diss_kappa_deltaG}. From here, we can see that incorporating $J_0(\Delta G)$ in $\kappa$ effectively captures the onset of instability that we observe experimentally. Physically, this dependence may be thought to arise from a
process where increasing state amplitudes within the cavities leads the amplifiers to introduce and amplify existing incoherent noise. This amplified noise directly competes with the intrinsic dissipation of each cavity, thereby reducing the overall effective dissipation rate. Later, we will see that this loss model, $\kappa_0(\Delta G)$, does not only accurately predict the onset of instability, but also explains the monotonic increase in transmission amplitude as a function of $\Delta G$ that we observed experimentally (\suppfigref{fig:f_phi_transmission}).

\subsection{Characterizing the dynamical matrix: Off-diagonal coupling coefficients}

\begin{figure}[!htb]
\centering
\includegraphics[width=.9\textwidth]{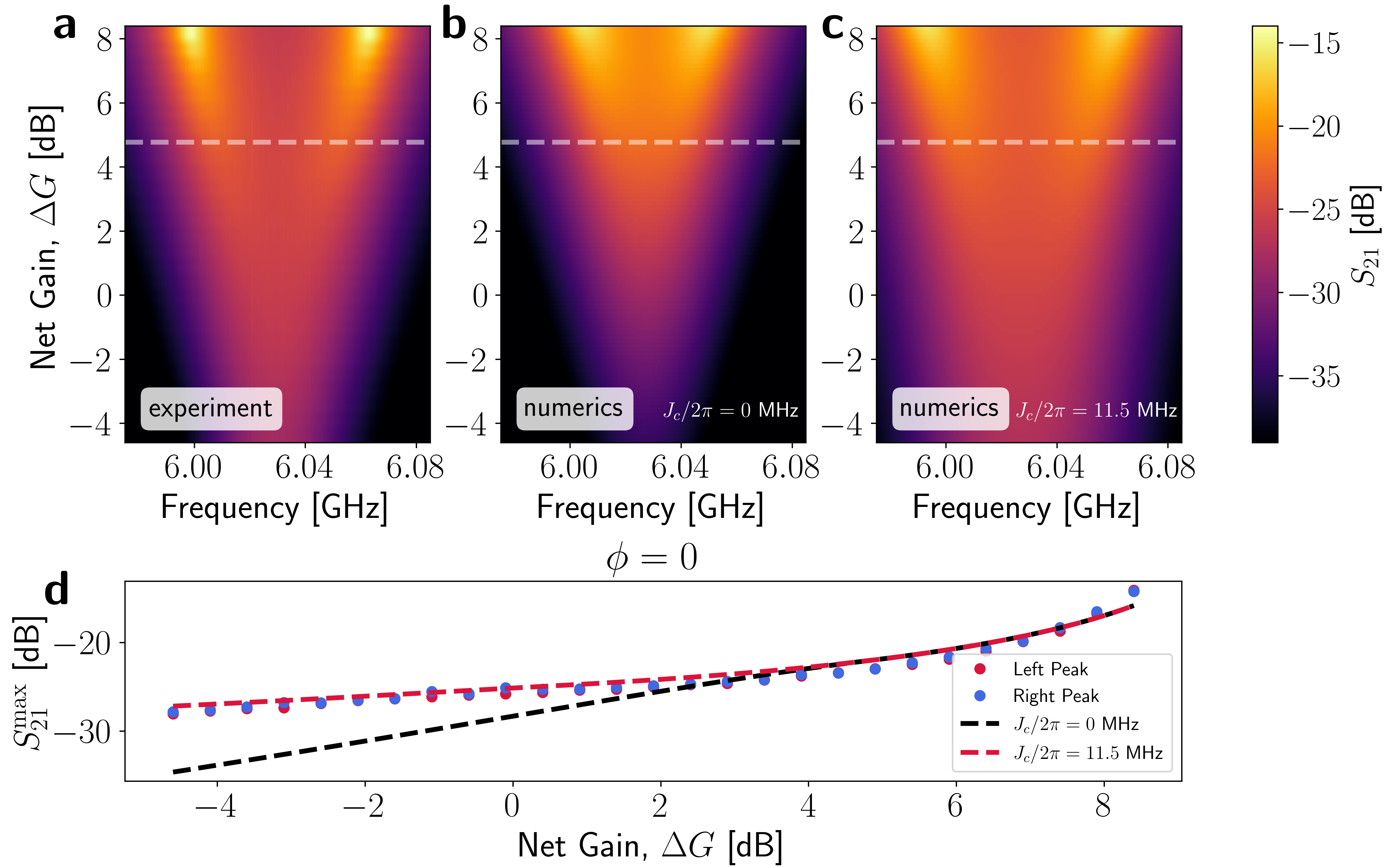}
\caption{\textbf{Fine-tuning weak-drive transmission features at $\mathbf{\phi = 0}$:} (\textbf{a}) Experimental and (\textbf{b-c}) simulated transmission spectra for the dimer considering \textbf{b} $J_c/2\pi= 0$ MHz, and \textbf{c} $J_c/2\pi= 11.5$ MHz. \textbf{d} Experimental (solid circles) and numerical (dashed lines) maximum transmission, $S_{21}^{\text{max}}$, as a function of $\Delta G$ for the left and right peak in the spectrum. The horizontal white dashed line in \textbf{a-c} represents the $\Delta G$ value at which the system becomes unstable at $\phi = \pi$. Experimental and numerical results in \textbf{a-d} consider an input drive power of $P_{d} = -30$ dBm.} 
\label{fig:f_phi_transmission}
\end{figure}

To better align our theoretical model with the experimental data, we next address a discrepancy observed in the onset of mode splitting at $\phi = 0$ based on the experimentally recorded transmission spectrum, $S_{21}$, depicted in \,\suppfigref{fig:f_phi_transmission}\textbf{a}. Theoretically, 
using the dynamical matrix in Eq.\,\eqref{eq:general_form}, mode splitting is predicted to occur at the {\em same} $\Delta G$ value for which the system becomes unstable at $\phi = \pi$, as depicted by the horizontal white dashed line in each of the plots in \,\suppfigref{fig:f_phi_transmission}\textbf{b}. 
Instead, transmission experiments show that splitting begins at a {\em significantly lower} $\Delta G$, suggesting a more substantial coherent-hopping effect at $\phi = 0$, which would then lead to earlier mode splitting.

Numerically, we can calculate the transmission spectra observed experimentally when we apply a sufficiently weak drive. In the 
stable regime the system converges towards a unique equilibrium point, which can be computed as
\begin{align}
\begin{bmatrix}
\alpha_1^\text{eq} \\ \alpha_2^\text{eq}
\end{bmatrix} = -A_0^{-1}(\omega_d,\Delta G,\phi) \begin{bmatrix}
\epsilon \\ 0 
\end{bmatrix}, 
\label{eq:transmission-theory}
\end{align} 
where we have now included the explicit dependence on the newly relevant drive frequency,  $\omega_d$. 
Mathematically, we can account for the above-mentioned coherent-hopping mechanism by introducing a function $f(\phi)$, which directly modifies the off-diagonal hopping coefficients and is defined as in the main text, namely:
\begin{align}
f(\phi) &= i J_c \cos\left(\frac{\phi}{2}\right) e^{i\phi/2}.
\end{align} 
Here, $J_c/2\pi = 11.5$ MHz represents the strength of this additional phenomenological coherent hopping. While the origin of $f(\phi)$ remains unclear, it likely involves the effect of constructive interference between the two modes, enhancing the effective hopping and resulting in the frequency splitting of the eigenmodes at lower $\Delta G$ values in \,\suppfigref{fig:f_phi_transmission}\textbf{b} ($J_c/2\pi = 0$ MHz) than we observe in \,\suppfigref{fig:f_phi_transmission}\textbf{c} ($J_c/2\pi = 11.5$ MHz). In this context, our updated dynamical matrix for the low-power, stable regime, which incorporates $f(\phi)$ and is henceforth referred to as the ``linear model", is governed by the dynamical matrix:
\begin{align}
A_0(\omega_d,\Delta G,\phi) = \begin{bmatrix}
-i(\omega_c-\omega_d)-\kappa(\Delta G) & [-iJ_{0}(\Delta G) - f(\phi)] e^{-i\phi} \\ -iJ_{0}(\Delta G) - f(\phi) & -i(\omega_c-\omega_d) - \kappa(\Delta G)
\end{bmatrix}.
\label{eq:linear_model_dynamical_matrix}
\end{align} 

Following Eq.\,\eqref{eq:transmission-theory}, the $S_{21}$ transmission spectra are calculated as the ratio of the measured output power of cavity two to the input drive power,  for obtaining the numerical result in \,\suppfigref{fig:f_phi_transmission}. That is: 
\begin{align}
    S_{21}(\omega_d,\Delta G,\phi) = \frac{P_\text{out}}{P_\text{in}} = \frac{\hbar \omega_d|\alpha_2^\text{eq}|^2 \kappa_\text{out}}{\hbar \omega_d \epsilon^2/\kappa_\text{in}} = \kappa_\text{in}\kappa_\text{out}\frac{|\alpha_2^{\text{eq}}|^2}{\epsilon^2},
\end{align} 
where $\alpha_{2}^{\textrm{eq}}$ is the DC-component of the unique asymptotic solution $\alpha_{2}(t)$ and the dependence on $\Delta G$ and $\phi$ is implicitly contained in $\alpha_2^\text{eq}$. Moreover, the conversion to dB is accomplished via $S_\text{21}$ [dB] $=10\log_\text{10}(S_{21})$. 

\suppfigref{fig:f_phi_transmission}\textbf{d} shows the experimental (solid circles) and numerical (dashed lines) maximum transmission, $S_{21}^{\text{max}}$, as a function of $\Delta G$. Experimentally, we observe a monotonic increase of $S_{21}^{\text{max}}$ with $\Delta G$, and the overall trend seems to be very well captured by the linear model with and without $f(\phi)$. \suppfigref{fig:f_phi_transmission}\textbf{d} shows that including the $f(\phi)$ term in Eq.\,\ref{eq:linear_model_dynamical_matrix} not only leads to earlier mode splitting, but also closely matches the observed $S_{21}^{\text{max}}$ at low $\Delta G$. This suggests an enhanced coherent hopping effect at $\phi = 0$, which increases $S_{21}^{\text{max}}$ at these lower $\Delta G$ values.

Overall, we emphasize that the inclusion of the $f(\phi)$ term significantly enhances the agreement between experiment and simulation while maintaining the stability characteristics of the system outlined in \,\suppfigref{fig:phase_diag}\textbf{a}. This approach provides a comprehensive and accurate representation of the dynamical matrix, capturing all experimental features of interest to a remarkable quantitative extent in the stable regime and for sufficiently weak drive. In the unstable regime or for stronger drives, however, we do need to consider the full nonlinearity in our model, as we will discuss next. 

\subsection{Model for nonlinearity}

Our experimental results indicate that when the system becomes unstable, a sudden and dramatic increase in the amplitude of the two coupled modes takes place. This 
drives the amplifiers into saturation, thereby compressing the gain in a power-dependent fashion. In turn, this steers the system towards a stable LC where the saturation characteristics of the amplifier determine the properties of the self-sustained mode. Consequently, to fully capture the dynamics observed, our model for the hopping function must incorporate a nonlinear dependence 
upon the field amplitudes, $|\alpha_i|^2$. The following continuous piece-wise function defines this relationship:
\begin{equation}
\label{eq:j_hop}
    J(\Delta G; |\alpha_i|^2) = \kappa_c 10^{\Delta G/20} f_{\textrm{G}}({\abs{\alpha}_{i}}^2),
\end{equation}
with
\begin{equation}
\label{eq:def_f_lG}
   f_{\textrm{G}}\big({\abs{\alpha}_{i}}^2\big) \equiv \begin{cases} 
 1 &, \text{if } |\alpha_i|^2 \leq |\alpha_\text{sat}|^2, \\
\frac{b_G + \hbar \omega_c |\alpha_{\text{sat}}|^2 \kappa_c}{b_G + \hbar \omega_c |\alpha_i|^2 \kappa_c} &, \text{if } |\alpha_i|^2 > |\alpha_\text{sat}|^2,
\end{cases} 
\end{equation}
where the parameters $b_G = 8.6$ mW and $|\alpha_{\text{sat}}|^2$ characterize the saturation behavior of the amplifier. Note that 
$|\alpha_{\text{sat}}|^2 = P_{\text{sat}} / \hbar \omega_c \kappa_c$, where $P_{\text{sat}} = 0.9981$ mW is the saturation power of the amplifiers. 
Additionally, for increased accuracy of the time-domain numerical analysis from the main text, we allow for the intra-cavity dissipation rates to be nominally different. The final values used in the numerical simulations can be found in \supptabref{tab:paramtab}.

At this point, we have introduced all of the terms required to define the final form of our EOMs, namely,
\begin{eqnarray}
\label{eq:dynamics_full}
\dot{\bm{\alpha}} &=& A(|\alpha_1|^2, |\alpha_2|^2) \bm{\alpha} + \epsilon\textbf{B},  
\end{eqnarray}
with \begin{align}
A({\abs{\alpha_{1}}}^2, {\abs{\alpha_{2}}}^2) = \begin{bmatrix}
-i (\omega_c - \omega_d) - \kappa_{1} (\Delta G, |\alpha_1|^2) & (-i J(\Delta G, |\alpha_2|^2) - f(\phi)) e^{-i\phi} \\
-i J(\Delta G, |\alpha_1|^2) - f(\phi) & -i (\omega_c - \omega_d) - \kappa_{2} (\Delta G, |\alpha_2|^2)
\end{bmatrix}, 
\label{eq:dynamical-matrix-f-phi-nl}
\end{align} where $\kappa_{1(2)}({\abs{\alpha_{i}}}^2 ; \Delta G) = 2(\kappa_{\text{int,1(2)}} + \kappa_{\text{in(out)}} + \kappa_c) - J(\Delta G, |\alpha_{1(2)}|^2)$, and $J(\Delta G, |\alpha_{1(2)}|^2)$ is defined in Eq.\,\eqref{eq:j_hop}. 

\subsection{
Parameters used in numerical simulations}

In \supptabref{tab:paramtab} we provide a summary of all the parameters used in our numerical simulations for solving the EOMs in
Eq.\,\eqref{eq:dynamical-matrix-f-phi-nl}. 
%
\begin{table}[!htb]
\begin{tabular}{c|c|c|c|c}
Parameter name & Symbol & Fixed/Tunable & Value/Range
\\
\hline \hline
Cavity 1 frequency & $\omega_1/2\pi$ & Fixed & $ 6.027(5)$ GHz
\\
Cavity 2 frequency & $\omega_2/2\pi$ & Fixed & $6.027(5)$ GHz
\\
Cavity 1 internal loss rate & $\kappa_{\text{int},1}/2\pi$ & Fixed & $ 4.1(1)$ MHz
\\
Cavity 2 internal loss rate & $\kappa_\text{int,2}/2\pi$ & Fixed &  $4.0(1)$ MHz
\\
Drive coupling rate & $\kappa_\text{in}/2\pi$ & Fixed & $ 2.5(2)$ MHz
\\
Readout coupling rate & $\kappa_\text{out}/2\pi$ & Fixed & $ 2.3(2)$ MHz
\\
Inter-cavity coupling rate & $\kappa_c/2\pi$ & Fixed & $ 8.7(1)$ MHz
\\
Additional coherent coupling & $J_c/2\pi$ & Fixed & $11.5$ MHz 
\\
Characteristic gain & $G_0$ & Fixed & $20.3(2)$ dB 
\\
Net Gain & $\Delta G$ & Tunable & $[-4.6, 8.4]$ dB 
\\
Net leg phase difference & $\phi$  & Tunable & $[0,2\pi]$ rad 
\\
Drive frequency & $\omega_d/2\pi$ & Tunable & $[5.975, 6.085]$ GHz 
\\
Drive strength & $P_d$ & Fixed &  $-30$ dBm 
\end{tabular}
\caption{All parameters of the model. The values specified herein are those directly utilized in the numerical solutions for $S_{21}$, onset of instability, LC solutions, and synchronization dynamics as presented in the main manuscript.}
\label{tab:paramtab}
\end{table}
Note that 
the parameters $\Delta G$ and $\phi$ that we tune in our theoretical model 
are actually not directly adjustable in experiments; they are instead derived from intensive calibrations of the components included in our device. These values are determined from the settings of two digital attenuators, $\Gamma_{1\to 2}$ and $\Gamma_{2\to 1}$, in dB, and the phase shifter $\phi_\text{exp}$, in degrees. This process, known as the \textit{hash map}, acts as a lookup table created from experimental data,  that establishes an effective map between the input settings in the devices to the model parameters $\Delta G$ and $\phi$, as explained in Sec.~\ref{sec:hashmap}. The uncertainties for the cavity-associated parameters listed in \supptabref{tab:paramtab} are derived from the variance in parameter estimates obtained by fitting the reflection spectrum, $S_{11}$, from each corresponding port. Details of this fitting process are provided in Sec.~\ref{sec:experimental_parameters_cavity}.


\section{Exact analysis of Hopf bifurcation and limit cycle properties for the undriven system}

\label{supsec:analytics_hopf}

In this section, we again assume that the linear dissipation rates of the two cavities are equal.
Correspondingly, for the full nonlinear model, we have that
\begin{equation}
\label{eq:restriction_equal_dissipation_rates}
    \kappa_{1}({\abs{\alpha}}^2 ; \Delta G) = \kappa_{2}({\abs{\alpha}}^2 ; \Delta G),
\end{equation}

\subsection{Normal mode description and linear stability analysis}

The first step in describing the supercritical Hopf bifurcation, and the resulting LC properties, is to diagonalize the linearized system around the vacuum (since it is not driven), described by a dynamical matrix $A_{0} \equiv A(0,0),$ with $A({|{\alpha_{1}}|}^2, {|{\alpha_{2}}|}^2)$ defined as
in Eq.\,\eqref{eq:dynamical-matrix-f-phi-nl}. This is obtained by introducing the normal-mode amplitudes
\begin{equation}
\label{eq:normal_mode_amplitudes}
    \beta_{\pm} \equiv \frac{\pm e^{i \frac{\phi}{2}} \alpha_{1} - \alpha_{2}}{\sqrt{2}}.
\end{equation}
Indeed, still at the linearized level, the equations of motion of the normal mode amplitudes decouple:
\begin{equation}
\label{eq:linearized_EOMs_normal_modes}
    \dot{\beta}_{\pm} = \left( -   \kappa_{\pm,0}(\phi, \Delta G)
                               - i \delta \omega_{\pm,0}(\phi, \Delta G) \right) \beta_{\pm}, 
\end{equation}
where we have defined
\begin{subequations}
\begin{align}
    \kappa_{\pm,0}(\phi,\Delta G) &\equiv \kappa_{0}(\Delta G)
                                    \mp J_{0}(\Delta G) \sin\left(\frac{\phi}{2}\right), \label{eq:def_kp0}\\
    \delta \omega_{\pm,0}(\phi,\Delta G) &\equiv \omega_{\textrm{c}} - \omega_{\textrm{d}}
                                    \pm (J_{0}(\Delta G) + J_{\textrm{c}}) \cos\left(\frac{\phi}{2}\right). \label{eq:def_dwp0}
\end{align}
\end{subequations}

Since $J_{0}(\Delta G) \geq 0$, we can see that the $\beta_{+}$-mode always has a lower dissipation rate than the $\beta_{-}$-mode. In this work, we will consider the case where the $\beta_{-}$-mode is stable for any phase $\phi$ in the linearized regime. This is the case when $\kappa_{0}(\Delta\;\!G) > 0$. We will then assume that $\beta_{-}(t) \equiv 0, \forall t$.
We then wish to study what dynamics the $\beta{+}$-mode undergoes when it becomes unstable, accounting for the full nonlinearity of the system. As it turns out, taking $\beta_{-}(t) \equiv 0, \forall t$ is a valid Ansatz, in the sense that it leads to a subset of exact solutions for the full nonlinear system as well, when solving the remaining EOMs for $\beta_{+}$. The center-manifold theorem (see e.g.\ Theorem 3.2.1.\ in~\cite{Holmes}) guarantees that the obtained class of solutions captures the relevant asymptotic behavior around the bifurcation point where the $\beta_{+}$-mode
becomes unstable. In practice, numerically, we see that the Ansatz $\beta_{-}(t) \equiv 0, \forall t$ is valid also for large amplitudes $\abs{\beta_{+}}$, as is the case e.g.\ for large limit cycle solutions. A discussion on possible explanations for this, as well as an analogous discussion for the driven model, is presented at the end of this section.

\subsection{Characterization of Hopf bifurcation}

Writing the full nonlinear system in Eq.\,\eqref{eq:dynamics_full} 
in the normal-mode basis, and putting $\beta_{-} = 0$, we obtain the (still exact) EOMs
\begin{equation}
\label{eq:full_EOMS_normal_modes}
\dot{\beta}_{+} = \left[- \kappa_{+}\big(\left|{\beta_{+}}\right|^{2} ; \phi, \Delta G\big)
                  - i   \delta \omega_{+}\big(\left|{\beta_{+}}\right|^{2} ; \phi,\Delta G\big) \right]
                  \beta_{+},
\end{equation}
with the relevant coefficients being given by
\begin{subequations}
\begin{align}
    \kappa_{+} \big(\left|{\beta_{+}}\right|^{2} ; \phi, \Delta G\big)
        &\equiv \kappa\bigg(\frac{{|\beta_{+}|}^2}{2} ; \Delta G \bigg)
            - \sin\left(\frac{\phi}{2}\right) J\bigg(\frac{{|\beta_{+}|}^2}{2}; \Delta G\bigg) \nonumber\\
        &= 2 \left( \kappa_{\textrm{int}} + \kappa_{\textrm{in/out}} + \kappa_{\textrm{c}} \right)
           - \bigg(\sin{\left(\frac{\phi}{2} \right)} + 1 \bigg)
                   J{\bigg(\frac{\left|{\beta_{+}}\right|^{2}}{2} ; \Delta G \bigg)},\\
    \delta \omega_{+}\big(\left|{\beta_{+}}\right|^{2} ; \phi, \Delta G\big)
        &= \omega_{\textrm{c}}-\omega_{\textrm{d}}
            - \bigg(J_{\textrm{c}} + J{\bigg(\frac{\left|{\beta_{+}}\right|^{2}}{2} ; \Delta G\bigg)}\bigg)
               \cos{\left(\frac{\phi}{2} \right)}.
\end{align}
\end{subequations}
Introducing polar coordinates $(\theta_{+}, R_{+})$ such that $\beta_{+} = i R_{+} e^{- i \theta_{+}},$
the EOMs then perfectly decouple,
\begin{subequations}
\begin{align}
    \dot{\theta}_{+} &= \delta \omega_{+}\left(R_{+}^{2} ; \phi, \Delta G\right),\\
    \dot{R}_{+}      &= - \kappa_{+}\left(R_{+}^{2} ; \phi, \Delta G\right) R_{+}.
\end{align}
\end{subequations}
Inspecting the $R_{+}$-dependence of the terms in $\delta \omega_{+}\left(R_{+}^{2} ; \phi, \Delta G\right)$
and $\kappa_{+}\left(R_{+}^{2} ; \phi, \Delta G\right)$,
we can identify both a Duffing-type nonlinearity, and a Van der Pol-type nonlinearity, respectively \cite{Holmes}. We can thus \textit{a priori} expect an amplitude-dependence of both the frequency and linewidth of the LC.

For the explicit analysis, it is instructive to consider the specific form in Eq.\,\eqref{eq:j_hop} for the nonlinear hopping function
$$J(R_{+}^2 / 2; \Delta G) = 10^{\Delta G / 20} \kappa_{\textrm{c}}
                             f_{\textrm{G}}\bigg(\frac{R_{+}^2}{2}\bigg),$$
where $f_{\textrm{G}}\left(\frac{R_{+}^2}{2}\right)$ is as defined in Eq.\,\eqref{eq:def_f_lG}.
To solve the radial EOM exactly in general, all we need is the fact that $f_{\textrm{G}}$
is a non-decreasing function, with
$f_{\textrm{G}}\Big( \big[0, {|\alpha_{\textrm{sat}}|}^2 \big]\Big) = \{1\}$ and $f_{\textrm{G}}$ is strictly monotonous and invertible on $({|\alpha_{\textrm{sat}}|}^2, \infty)$, with corresponding image $(0,1)$. Going forward, we will write $f_{\textrm{G}}^{-1}$ without further ado,
when its argument is contained in $(0,1)$.
It is then easy to see that $\kappa_{+}(R_{\textrm{LC}}^2; \phi, \Delta G) = 0$ has a non-zero solution $R_{\textrm{LC}} > 0$ if and only if the origin is unstable; i.e., when
$$\kappa_{+,0}(\phi;\Delta G) = \kappa_{0}(\Delta\;\!G) - J_{0}(\Delta\;\!G) \sin{\left(\frac{\phi}{2} \right)} < 0,$$ 
in which case the unique solution in question is given by
\begin{equation}
\label{LC_amplitude_abstract}
    R_{\textrm{LC}}^2 = 2 f_{\textrm{G}}^{-1}\!\left(\frac{ 2 ( \kappa_{\textrm{int}}+\kappa_{\textrm{in/out}}+\kappa_{\textrm{c}})}{10^{\Delta G / 20} \kappa_{\textrm{c}} \left(\sin{\left(\frac{\phi}{2} \right)} + 1\right)}\right).
\end{equation}
Note that the argument of $f_{\textrm{G}}^{-1}$ is indeed contained in $(0,1)$, given the condition for vacuum instability. If we now work this out, we obtain the explicit expression reported in Eq.\,(10) of the main text,
\begin{equation}
\label{eq:limit_cycle_amplitude}
    n_{\textrm{LC}} \equiv R^{2}_{LC} = 
    \frac{10^{\Delta G / 20} \left(1 + \sin{\left(\frac{\phi}{2} \right)}\right) \left(\kappa_{\textrm{c}}^{2} \hbar \omega_{\textrm{c}}  {|\alpha_{\textrm{sat}}|}^2 + \kappa_{\textrm{c}} b_{\textrm{G}}\right) - 2 b_{\textrm{G}} \left(\kappa_{\textrm{c}} + \kappa_{\textrm{in/out}} + \kappa_{\textrm{int}}\right)}{2 \kappa_{\textrm{c}} \omega_{\textrm{c}} \hbar \left(\kappa_{\textrm{c}} + \kappa_{\textrm{in/out}} + \kappa_{\textrm{int}}\right)} .
\end{equation}

This expression is only valid in the vacuum-unstable regime, in which case it is straightforward to verify that indeed the LC amplitude immediately jumps past the saturation amplitude:
$n_{\textrm{LC}} > {\abs{\alpha_{\textrm{sat}}}}^2$.
The underlying physical reason for this is that the function $f_{\textrm{G}}$ is perfectly constant in $[0, {\abs{\alpha_{\textrm{sat}}}}^2]$,
whereas for a ``textbook'' Hopf bifurcation, its second-order Taylor series around ${\abs{\alpha}}^2 = 0$ does not vanish (and has a negative sign, for the supercritical case).
However, the fact that $f_{\textrm{G}}$ here is perfectly flat until ${\abs{\alpha}}^2 > {|\alpha_{\textrm{sat}}|}^2$ (and hence the system is still linear there)
and subsequently \emph{decreases}, implies that we can  straightforwardly ``shift'' the analysis of the Hopf bifurcation to ${\abs{\alpha}}^2 > {|\alpha_{\textrm{sat}}|}^2$. Moreover, we note that while the model is phenomenological and does not explicitly include noise, its agreement with experiment suggests that it captures the key physical effects.

\subsection{Derivation of the remaining limit cycle properties}

We can re-insert the found value for $R_{\textrm{LC}}$ into the angular EOM, to find
the angular frequency of the LC.  It is instructive to keep the more general
expression~\eqref{LC_amplitude_abstract} for $R_{LC}^{2}$ to do this. Working this out step by step, we find
\begin{subequations}
\begin{align}
    \delta \omega_{\textrm{LC}} \equiv \delta \omega_{+}\left(R_{\textrm{LC}}^{2} ; \phi, \Delta G\right)
        & = \omega_{\textrm{c}}-\omega_{\textrm{d}} - \left(J_{\textrm{c}} + J{\left(\frac{R_{\textrm{LC}}^{2}}{2} \right)}\right) \cos{\left(\frac{\phi}{2} \right)}\\
        & = \omega_{\textrm{c}}-\omega_{\textrm{d}} - \left(\frac{2 (\kappa_{\textrm{int}}+\kappa_{\textrm{in/out}}+\kappa_{\textrm{c}})}{\sin{\left(\frac{\phi}{2} \right)} + 1} + J_{\textrm{c}}\right) \cos{\left(\frac{\phi}{2}
        \right)}\label{eq:LC_frequency}
\end{align}
\end{subequations}
Note that the exact functional dependence of $f_{\textrm{G}}$ does not change the value of the LC frequency.
This stems from the fact that the exact same amplifier saturation function $f_{\textrm{G}}$ determines the nonlinear part of {\em both} the radial and angular EOMs in~\eqref{eq:full_EOMS_normal_modes}.
For the same reason, the LC amplitude does not depend on the symmetric net gain $\Delta G$, but only on $\phi$ in terms of tunable parameters. As a consequence, the frequency of the LC can be obtained by considering the frequency of the unstable normal mode at the threshold of instability (eliminating the variable $\Delta G $ in the process). Indeed, if we substitute 
\[10^{\Delta  G / 20} = \frac{2 \left(\kappa_{\textrm{c}} + \kappa_{\textrm{in/out}} + \kappa_{\textrm{int}}\right)}{\kappa_{\textrm{c}} \left(\sin{\left(\frac{\phi}{2} \right)} + 1\right)}\]
into Eq.\,\eqref{eq:def_dwp0}, we consistently obtain Eq.\,\eqref{eq:LC_frequency}.

To calculate the local rate of (exponential) convergence towards the LC, we linearize the radial EOM around $R = R_{LC}$. Defining $\delta R_{+} \equiv R_{+} - R_{LC}$
and Taylor-expanding up to second order in $\delta R_{+}$, we obtain
\begin{align}
    \delta \dot{R}_{+} = - R_{\textrm{LC}} \, \kappa_{\textrm{+}}{\left(R_{\textrm{LC}}^{2}; \phi,\Delta\!G \right)}
    - \left [2 R_{\textrm{LC}}^{2} \, \kappa_{\textrm{+}}{'}(R_{\textrm{LC}}^{2} ; \phi,\Delta\!G)
    + \kappa_{\textrm{+}}{\left(R_{\textrm{LC}}^{2} ; \phi,\Delta\!G \right)}\right] \delta\!R_{+}
     + \mathcal{O}\left(\delta\!R_{+}^{2}\right).
\end{align}
Using the fact that at the LC solution, $\kappa_{\textrm{+}}{\left(R_{\textrm{LC}}^{2}; \phi,\Delta\!G \right)} = 0$ by definition, we get the linearized EOM for the perturbation $\delta R_{+}$,
\begin{equation}
    \delta \dot{R}_{+} \simeq - \kappa_{LC}(\phi , \Delta G) \delta R_{+},
\end{equation}
where we have defined the local convergence rate
$$\kappa_{LC}(\phi , \Delta G) \equiv 2 R_{\textrm{LC}}^{2} \, \kappa_{\textrm{+}}{'}(R_{\textrm{LC}}^{2} ; \phi,\Delta\!G).$$

Now all there is left for us to do is to work out the specific parameter dependence of $\kappa_{LC}(\phi, \Delta G)$.
For this it is instructive to first realize that for our specific nonlinearity, we have that
\begin{equation}
    f_{G}'(R_{+}^2 / 2) = - \frac{\kappa_{\textrm{c}} \omega_{\textrm{c}} \hbar}{\frac{R_{+}^{2} \kappa_{\textrm{c}} \omega_{\textrm{c}} \hbar}{2} + b_{\textrm{G}}} f_{\textrm{G}}{\left(\frac{R_{+}^{2}}{2} \right)}.
\end{equation}
This allows us to write 
\begin{equation}
    \kappa_{LC}(\phi ; \Delta G) = 2 R_{\textrm{LC}}^{2} \left(2 (\kappa_{\textrm{int}}+\kappa_{\textrm{in/out}}+\kappa_{\textrm{c}}) + \frac{10^{\Delta G / 20} \kappa_{\textrm{c}}^{2} \omega_{\textrm{c}} \hbar \left(\sin{\left(\frac{\phi}{2} \right)} + 1\right) f_{\textrm{G}}{\left(\frac{R_{\textrm{LC}}^{2}}{2} \right)}}{\frac{R_{\textrm{LC}}^{2} \kappa_{\textrm{c}} \omega_{\textrm{c}} \hbar}{2} + b_{\textrm{G}}}\right),
\end{equation}
which, using the relation
\begin{equation}
    f_{G}(R_{LC}^2 / 2) = \frac{2 (\kappa_{\textrm{int}}+\kappa_{\textrm{in/out}}+\kappa_{\textrm{c}})}{10^{\Delta G / 20} \kappa_{\textrm{c}} \left(\sin{\left(\frac{\phi}{2} \right)} + 1\right)},
\end{equation}
can be rewritten as
\begin{equation}
   \kappa_{LC}(\phi ; \Delta G) = \frac{4 (\kappa_{\textrm{int}}+\kappa_{\textrm{in/out}}+\kappa_{\textrm{c}}) R_{\textrm{LC}}^{2} \left(R_{\textrm{LC}}^{2} \kappa_{\textrm{c}} \omega_{\textrm{c}} \hbar + 2 \kappa_{\textrm{c}} \omega_{\textrm{c}} \hbar + 2 b_{\textrm{G}}\right)}{R_{\textrm{LC}}^{2} \kappa_{\textrm{c}} \omega_{\textrm{c}} \hbar + 2 b_{\textrm{G}}}.
\end{equation}
Finally substituting the value for $R_{LC}$ found in Eq.\,\eqref{eq:limit_cycle_amplitude}, we obtain
\begin{equation}
\label{eq:dissipation_rate_LC}
   \kappa_{LC}(\phi ; \Delta G) = \frac{4 \left[10^{\Delta G / 20} \kappa_{\textrm{c}} \left(\kappa_{\textrm{c}} \omega_{\textrm{c}} \hbar {|\alpha_{\textrm{sat}}|}^2 + b_{\textrm{G}}\right) \left(\sin{\left(\frac{\phi}{2} \right)} + 1\right) - 2
   b_{\textrm{G}} \left(\kappa_{\textrm{c}} + \kappa_{\textrm{in/out}} + \kappa_{\textrm{int}}\right)\right]\left(\kappa_{\textrm{c}} + \kappa_{\textrm{in/out}} + \kappa_{\textrm{int}}\right)}{10^{\Delta G / 20} \kappa_{\textrm{c}}
   \left(\kappa_{\textrm{c}} \omega_{\textrm{c}} \hbar {|\alpha_{\textrm{sat}}|}^2 + b_{\textrm{G}}\right) \left(\sin{\left(\frac{\phi}{2} \right)} + 1\right)}.
\end{equation}

\subsection{Discussion on absence of other stable limit sets}\label{ssec:absence_other_limit_sets}

For the full nonlinear model, defined in~\eqref{eq:dynamics_full},\eqref{eq:dynamical-matrix-f-phi-nl} (or also Eqs. (8) or (9) in the main text),
we numerically observe that for all relevant parameter values, there is a unique asymptotic limit set towards which all trajectories converge, regardless of the initial condition of the system.
The limit set is either a stable equilibrium point, or a stable limit cycle, when the former equilibrium point is unstable.
In this section, we have described this limit set exactly analytically for the case where there is no driving,
and the uniqueness of the asymptotic behavior is seen to be guaranteed for two reasons. First, the stability of the $\beta_{-}$ mode
allows to reduce the system to a two-dimensional system (real) dynamical system, encoded in the real and imaginary parts of the $\beta_{+}$ mode amplitude.
This immediately excludes any chaotic behavior, since there is no explicit time-dependence.
Second, the monotonicity of the gain profile function $f_{\textrm{G}}$ was leveraged to show that there is always a unique steady state for the radial EOM for the $\beta_{+}$ - mode $-$ which decoupled from the angular EOM owing to the $U(1)$-symmetry.

In the driven case, no explicit analytical expression can be obtained for the solutions (also not in the long-time asymptotic regime),
while numerically we still observe that all initial conditions either converge towards towards a unique stable equilibrium point,
or to a unique globally attractive limit cycle (when again the equilibrium point loses its stability).
While there is no a priori reason for this to be the case, given that our system can be cast into a four-dimensional nonlinear system in terms of real quadratures,
and we don't have a rigorous theoretical argument providing an explanation, we can share an intuitive partial explanation.

We suspect that also in the driven case, while the stable manifold tangent to the (displaced, now) $\beta_{-}$ plane is only guaranteed to locally exist (see the Hartman-Grobman linearization or center manifold theorem,~\cite{Holmes}), for our system, it might be globally defined, or at least in a large enough domain for experimentally relevant initial conditions.
In this case, one can again reduce the system to a 2D manifold, where no chaos can occur.
We would like to stress that this is owing to the fact that the explicit time-dependence of the drive can be  exactly transformed away in the rotating frame, in the original model.

\subsection{Experimental characterization of the limit cycle}

\begin{figure}[htbp!]
\centering
\includegraphics[width=0.7\textwidth]{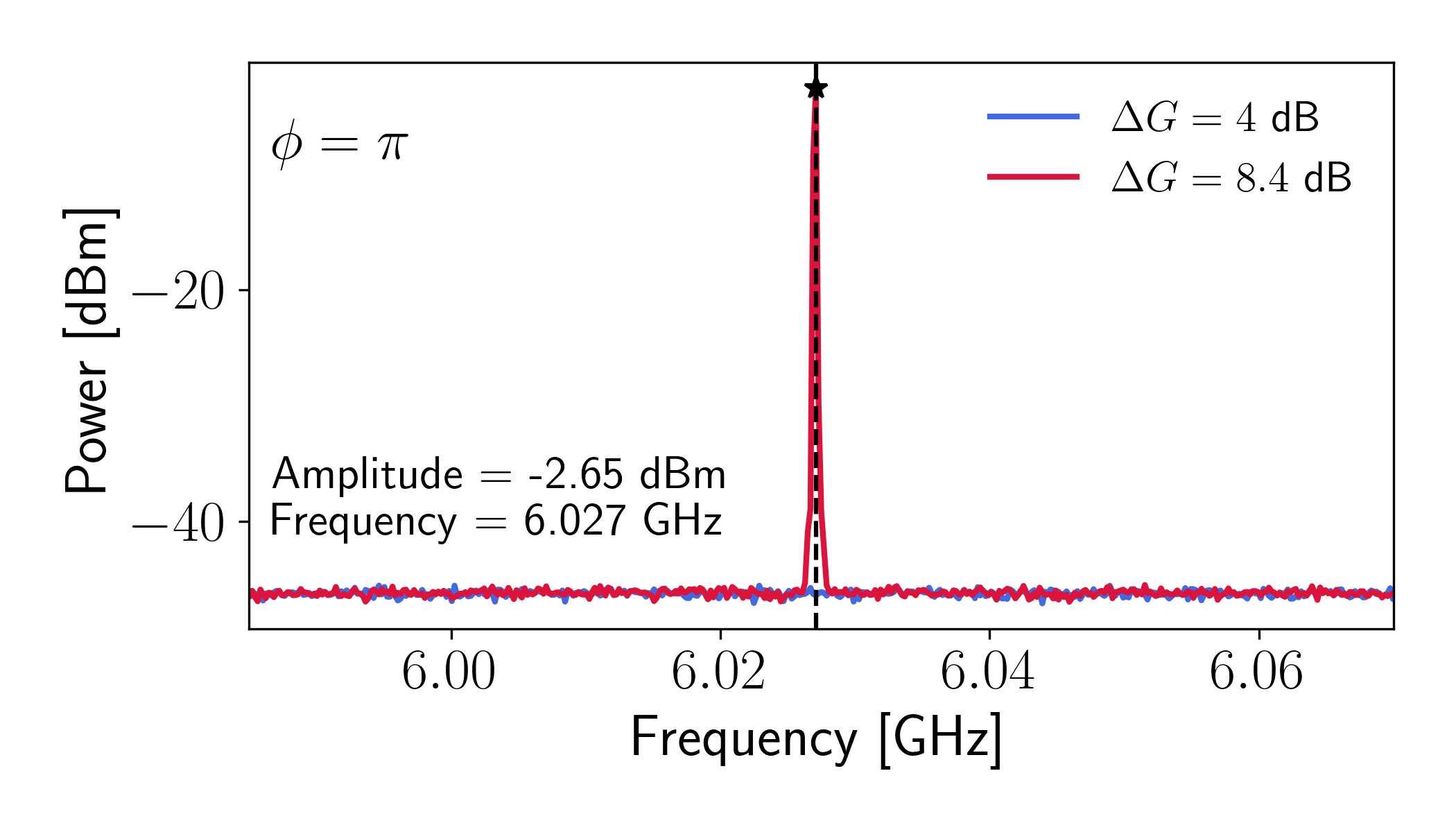}
\caption{\textbf{Experimental characterization of limit-cycle amplitude and frequency.} Emission spectra are shown for the dimer at two representative gain values, $\Delta G = 4.0$ dB (below threshold) and 8.4 dB (above threshold), with $\phi = \pi$. The limit cycle amplitude (black star) and frequency (black dashed line) are extracted from the spectra and plotted as a function of $\Delta G$ and $\phi$ to obtain the phase diagram shown in Fig. 3 \textbf{a,d} in the main manuscript.}

\label{fig:experimental_LC}
\end{figure}

We extract the limit cycle (LC) amplitude and frequency from experimental emission spectra using a simple peak-finding method. For each spectrum, we identify the frequency point with the maximum power above a fixed threshold of -44 dBm. The corresponding power and frequency define the LC amplitude and $\omega_{\text{LC}}$, respectively. The threshold of -44 dBm was chosen to robustly reject spurious detections arising from the baseline noise floor, particularly in regimes where no self-sustained oscillations are present. In practice, we find that this criterion is sufficient to reliably distinguish between noise and genuine LC signals across the entire parameter space explored in this study.

An example of this extraction procedure is shown in  \suppfigref{fig:experimental_LC}, where we plot emission spectra at $\phi = \pi$ for two different gain values: one below threshold ($\Delta G = 4.0$ dB) and one above threshold ($\Delta G = 8.4$ dB). The LC amplitude and frequency are marked with a black star and a vertical dashed line, respectively, for the above-threshold case. No LC peak is detected for the sub-threshold trace, as expected.

A similar procedure is applied to the numerically simulated emission spectra by identifying the dominant peak in their Fourier transforms. To suppress transient artifacts, we exclude the initial 20\% of the time trace, and we discard cases where $|\alpha_2|^2 < 0.001\%$, $|\alpha_{\text{sat}}|^2$ to avoid spurious results associated with vacuum solutions. Overall, the extracted amplitudes and frequencies are used to generate the phase diagrams shown in Figs. 3\textbf{a–b} and 3\textbf{d–e} of the main text. For the analytical model, $\omega_{\text{LC}}$ is obtained directly, without the need for further processing.

\section{Understanding the synchronization phenomena: phase locking}

\begin{figure}[htbp!]
\centering
\includegraphics[width=.8\textwidth]{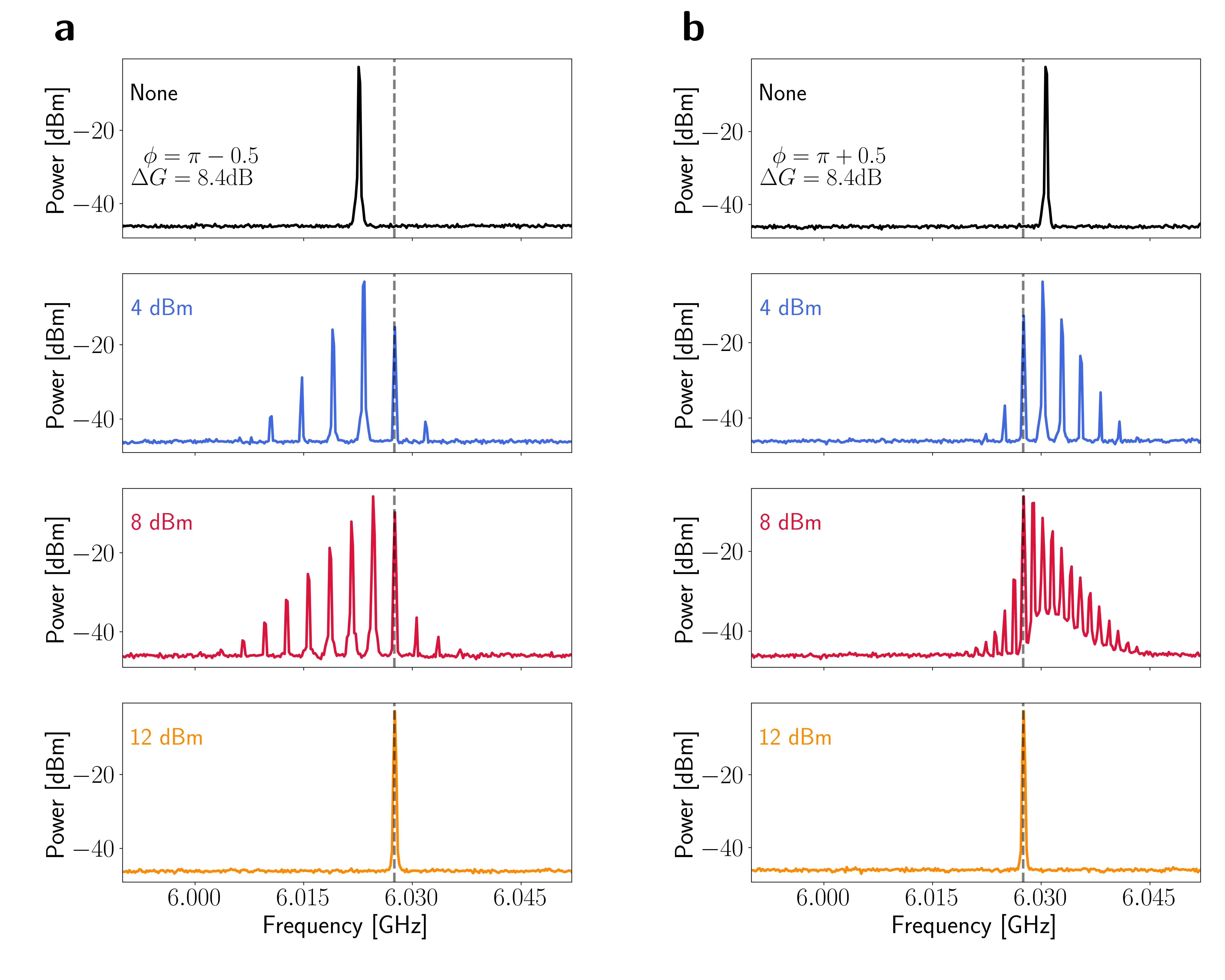}
\caption{\textbf{Experimental emission spectra of the driven and undriven dimer.} Measured emission spectra for the dimer at (\textbf{a}) $\phi = \pi - 0.5$ and (\textbf{b}) $\phi = \pi + 0.5$, with $\Delta G = 8.4$ dB. Spectra are shown for the undriven case and under driving at $\omega_d/2\pi = 6.027$ GHz (black dashed line), for drive powers $P_d = 4$, 8, and 12 dBm, respectively.}
\label{fig:phase_locking_sync_exp}
\end{figure}

To distinguish between phase locking and suppression of natural dynamics, we plotted the  emission spectra for the dimer set at $\Delta G = 8.4$ dB and symmetrically detuned phases $\phi = \pi \pm 0.5$. Emission spectra were recorded for the undriven case and for drive powers $P_d = 4$, 8, and 12 dBm at a fixed drive frequency $\omega_d/2\pi = 6.027$ GHz.

\begin{figure}[htbp!]
\centering
\includegraphics[width=.8\textwidth]{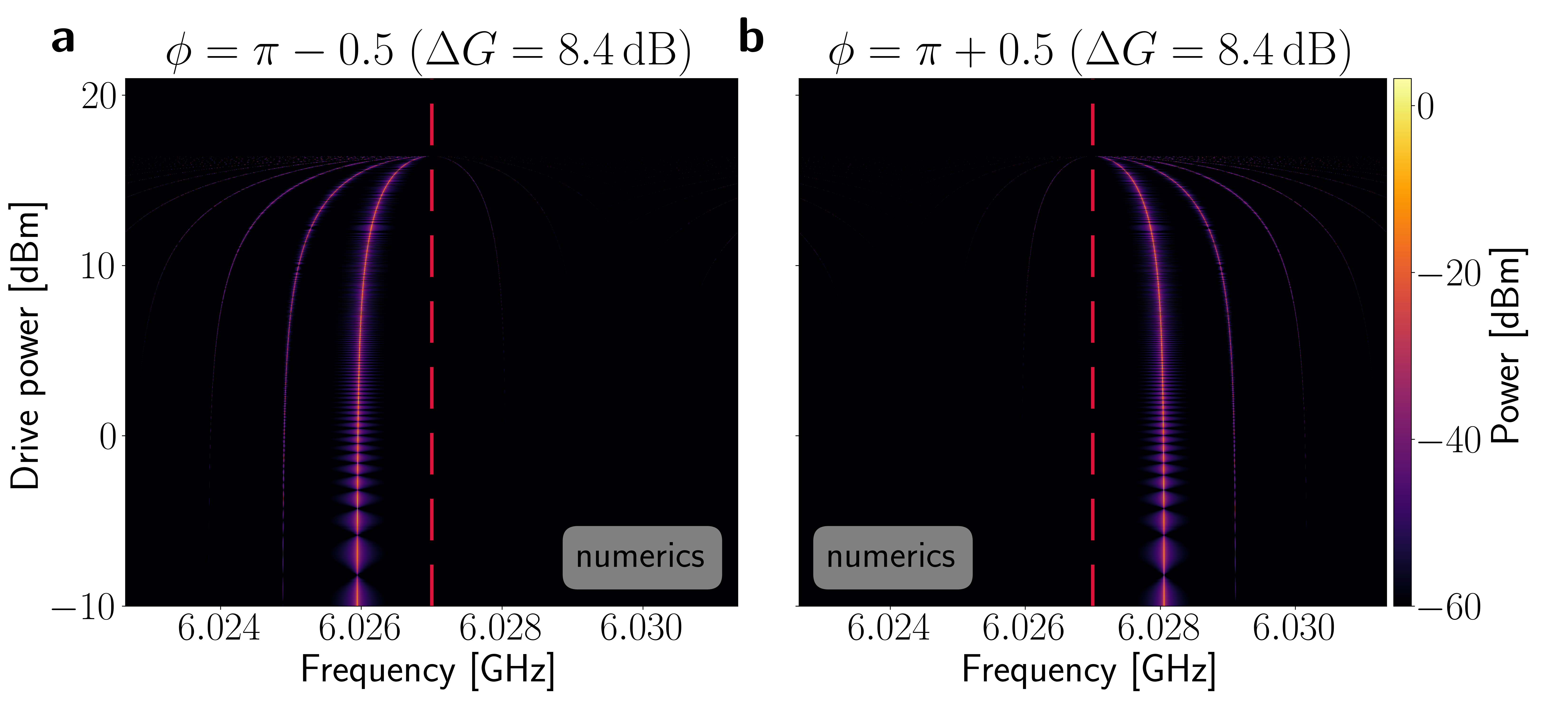}
\caption{\textbf{Simulated emission spectra for the driven dimer.} Simulated emission spectra for the dimer at (\textbf{a}) $\phi = \pi - 0.5$ and (\textbf{b}) $\phi = \pi + 0.5$, with $\Delta G = 8.4$ dB. An external drive at $\omega_d/2\pi = 6.027$ GHz (red dashed line) is applied, and the emission is characterized as the drive power $P_d$ is increased.}
\label{fig:phase_locking_sync_num}
\end{figure}

As shown in  \suppfigref{fig:phase_locking_sync_exp}, the measured spectra remain approximately symmetric about $\phi = \pi$ and reveal a progressive shift of the dominant spectral peak toward the drive frequency as $P_d$ increases. The main limit cycle peak in the undriven case, along with the higher-order harmonics generated under driving, coalesce into a single tone at $\omega_d$. This behavior is a hallmark of phase locking and differs qualitatively from the expected response in the suppression of natural dynamics, where the original peak would instead remain stationary and diminish in amplitude with increasing drive \cite{Balanov2009}.

To complement these measurements, we carried out numerical simulations under the same conditions. The results, shown in  \suppfigref{fig:phase_locking_sync_num}, exhibit excellent agreement with the experimental data, both in spectral structure and power dependence. Together, these findings confirm the presence of phase-locking synchronization in this regime and validate our modeling framework.

\begin{figure}[!tb]
\centering
\includegraphics[width=.55\textwidth]{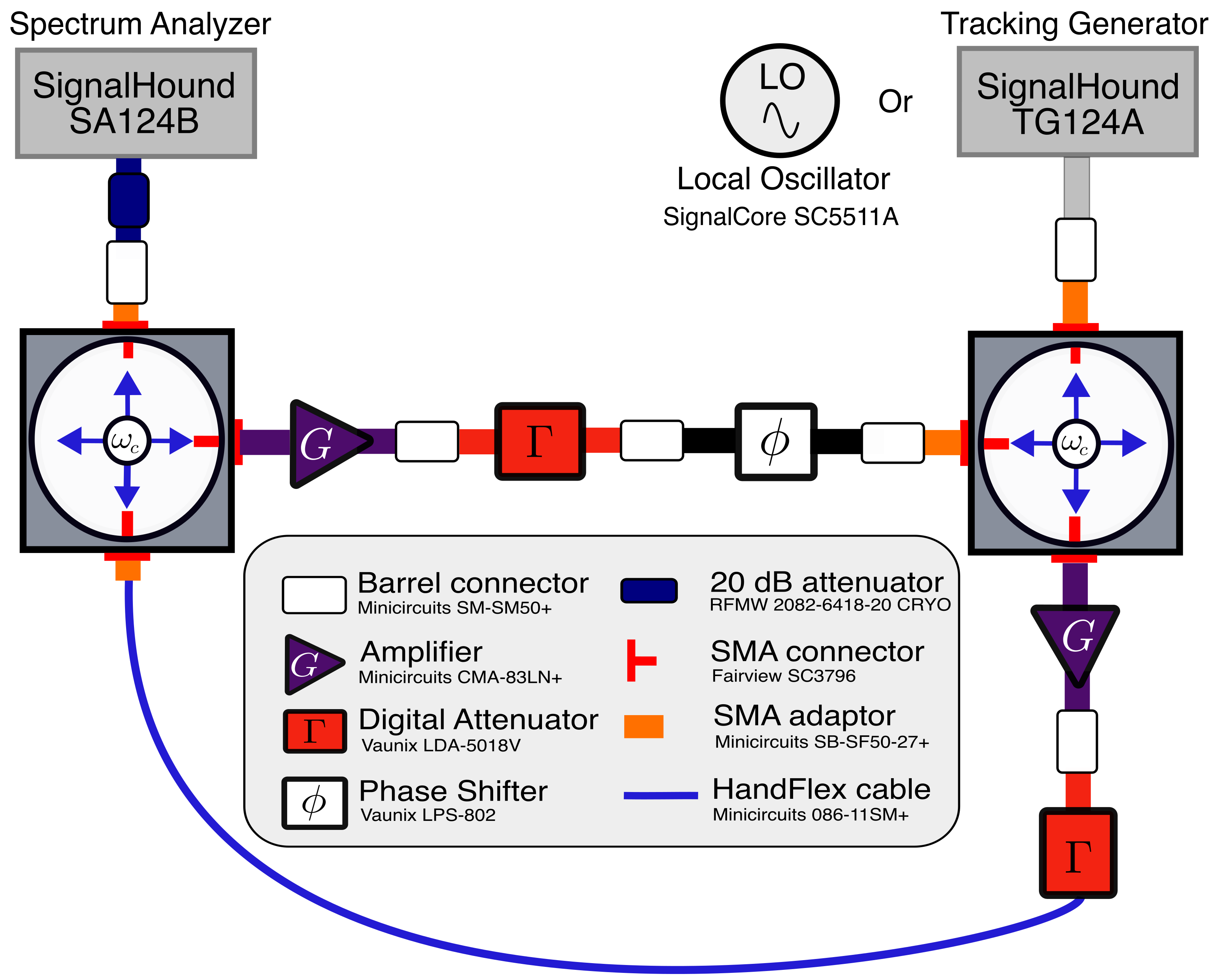}
\caption{\textbf{Schematic for our tunable, non-Hermitian, nonlinear microwave dimer.} This diagram illustrates the closed-loop configuration of two coupled oscillators (cavity 1 on the right and cavity 2 on the left), incorporating active (amplifiers) and passive (attenuators, phase shifters, adaptors, cables, connectors) elements to engineer the amplitude and phase of photon hopping. Each oscillator has three SMA connectors for coupling, driving, and readout functions. A synchronized tracking generator and spectrum analyzer facilitate transmission experiments across calibrated baseline spectra at constant drive strength. An external 20 dB attenuator is employed to protect the spectrum analyzer. Additionally, we utilize a local oscillator to probe the synchronization physics explained in the Main Text.}
\label{fig:device-schematic}
\end{figure}

\section{Device Schematics}

\suppfigref{fig:device-schematic} presents a schematic of our dimer in terms of components. The setup includes a tracking generator (SignalHound TG124A), synchronized with a spectrum analyzer (SignalHound SA124B), which serves as a scalar network analyzer and is utilized for collecting the data displayed in \,\suppfigref{fig:f_phi_transmission}\textbf{a}. The dimer comprises two microwave cavity resonators with mechanically tunable frequency and coupling rates. Each cavity is directly connected to an amplifier (Minicircuits CMA-83LN+), followed by a digital attenuator (Vaunix LDA-5018V) for tunable gain control. A phase shifter (Vaunix LPS-802) modifies the phase for the hopping path from cavity 2 to cavity 1, while a HandFlex cable (Minicircuits 086-11SM+) completes the reciprocal path, effectively closing the loop. SMA adapters (Minicircuits SM-SM50+) bridge connections between components where necessary. All components are designed to be 50 $\Omega$, ensuring impedance matching throughout the system in linear operation. An additional 20 dB attenuator protects the spectrum analyzer from high output powers. For LC experiments without an external drive, as shown in \,\suppfigref{fig:phase_diag}, a 50 $\Omega$ terminator is connected to the driving port of the second cavity. Additionally, when a drive of precise strength and frequency is required, an external signal generator (SignalCore SC5511A) is utilized.

\section{Methodology for Obtaining Experimentally Measured Parameters}
\label{sec:experimental_parameters_cavity}

This section provides a detailed overview of the experimental calibration process for our oscillators, particularly their resonance frequencies, $\omega_1$ and $\omega_2$, and the dissipation rates -- internal ($\kappa_{\text{int}}$), external ($\kappa_{\text{in/out}}$), and coupling ($\kappa_c$) rates for each oscillator, respectively.

The process starts by establishing a reflection measurement setup using a circulator (QCY-G0400801) to collect the reflection spectrum, $S_{11}$, which allows for the extraction of specific data from each individual port. We then adjust the machined SMA-connectors to achieve the desired coupling strength for each port. Additionally, we use a general fitting function formulated for a Hanger-type $\lambda/4$ resonator \cite{S11_fit, S11_fit_transfer_matrix} to determine the dissipation rates and resonance frequencies.
\begin{equation}
\label{eq:s11_fit}
    \left| S_{11} (\omega_{\text{probe}})\right| = - \left| \frac{\frac{\tilde{\omega}_c}{2Q_c}}{i(\tilde{\omega}_c - \omega_{\text{probe}}) + \tilde{\omega}_c \big(\frac{1}{Q_{\text{int}}} + \frac{1}{Q_c}\big)}\right| + S_{11,\text{baseline}}, 
\end{equation} 
where
\begin{align}
\tilde{\kappa}_{\text{int}} = \frac{\tilde{\omega}_c}{Q_{\text{int}}}, \qquad
\tilde{\kappa}_c = \frac{\tilde{\omega}_c}{2Q_c}.
\end{align} 
Hence, $\tilde{\omega}_c/2\pi$, $\tilde{\kappa}_c/2\pi$ and $\tilde{\kappa}_{\text{int}}/2\pi$ can be readily obtained as fitting parameters from the reflection spectrum of each port. Here, the \textit{tilde} notation represent the fitted resonance frequency, internal, and coupling dissipation rates obtained when interrogating each port individually. For instance, when obtaining $\tilde{\kappa}_c$ for the drive port, this fitting parameter will tell how much coupled the drive SMA pin is to the mode, namely $\kappa_{\text{in}}/2\pi$ in our model notation. A similar mapping is used to determine the other dissipation rates, specifically $\kappa_{\text{out}}/2\pi$ and $\kappa_c/2\pi$ for the readout and coupling ports, respectively. To avoid confusion, the final results are reported in \supptabref{tab:diss_rates_1}, and \supptabref{tab:diss_rates_2} following our regular model notation. With this iterative process, we carefully adjust each coupling rate to closely approximate a symmetric system, ensuring a close match for each corresponding port across the two cavities. Moreover, increasing the coupling rates through the adjustable couplers is known to introduce a loading effect, primarily impacting the resonant frequency of each oscillator. Therefore, this meticulous iterative process is designed to balance the loading effects from the coupling mechanisms, ensuring that the resonant frequency of the cavity remains as close as possible to the target value in every case.

\begin{figure}[!htb]
\centering
\includegraphics[width=.8\textwidth]{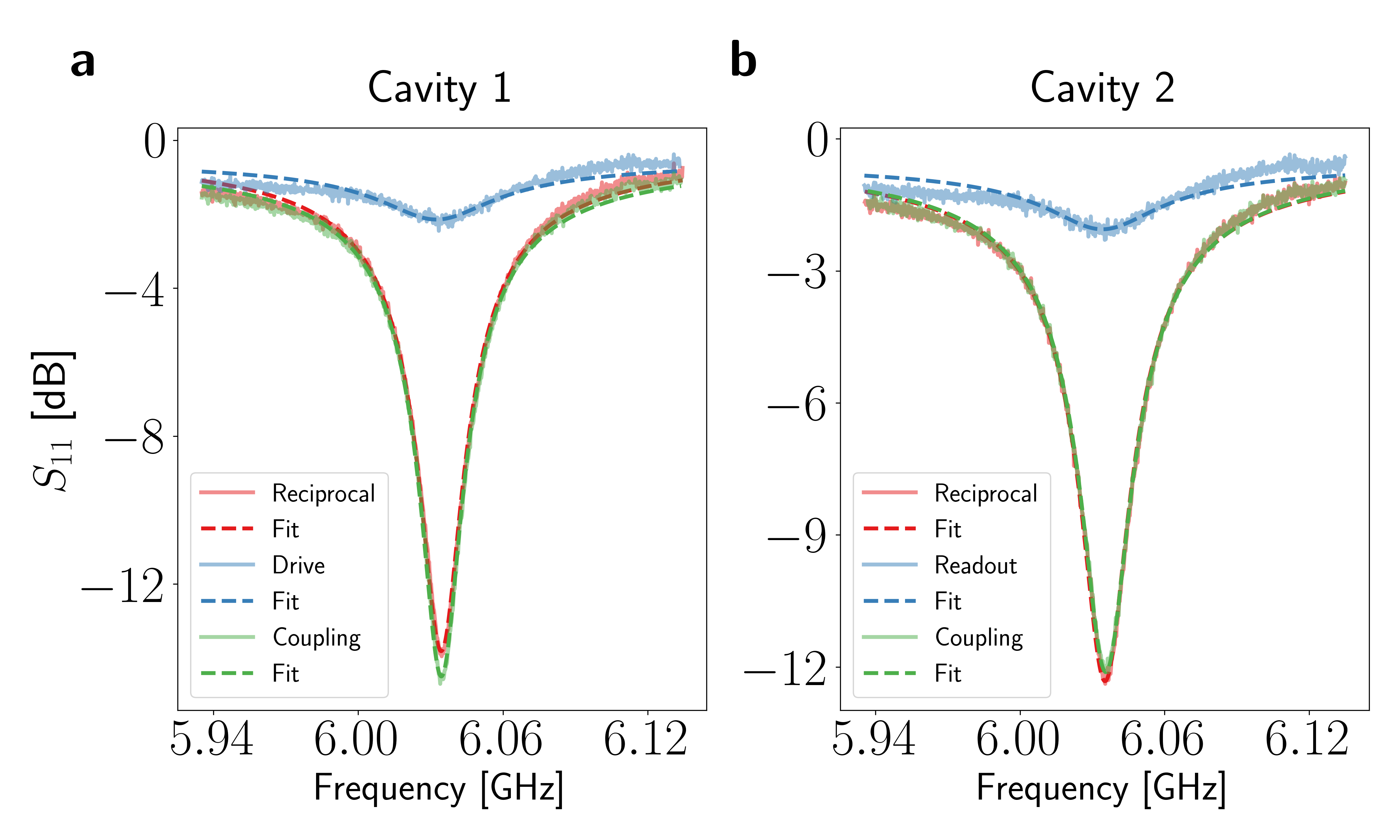}
\caption{\textbf{Calibrated reflection spectrum for each oscillator:} \textbf{a,} Drive cavity. \textbf{b,} Readout cavity. Each plot displays a broad, less prominent reflection dip from the under-coupled SMA pin, representing the drive and readout ports, respectively.  More pronounced dips correspond to the reciprocal and coupling ports in each cavity. All reflection spectra are centered around the target resonance frequency. Fitting parameters are detailed in \supptabref{tab:diss_rates_1} and \supptabref{tab:diss_rates_2}, respectively.}
\label{fig:dissipation-rates}
\end{figure}

\begin{table}[!htb]
    \centering
    \begin{minipage}{0.37\textwidth}
        \centering
        \renewcommand{\arraystretch}{1.5}
        \begin{tabularx}{\textwidth}{|l|X|}
            \hline
            \textbf{Port} & \textbf{Results} \\
            \hline
            Amplifier Port &
            $\begin{array}{l}
            \omega_1/2\pi = 6.0340(5) \, \text{GHz} \\
            \kappa_c/2\pi = 9.9(1) \, \text{MHz} \\
            \kappa_{\text{int, 1}}/2\pi = 5.7(1) \, \text{MHz}
            \end{array}$ \\
            \hline
            Reciprocal Port &
            $\begin{array}{l}
            \omega_1/2\pi = 6.0340(6) \, \text{GHz} \\
            \kappa_c/2\pi = 10.1(1) \, \text{MHz} \\
            \kappa_{\text{int, 1}}/2\pi = 5.7(1) \, \text{MHz}
            \end{array}$ \\
            \hline
            Drive Port &
            $\begin{array}{l}
            \omega_1/2\pi = 6.032(5) \, \text{GHz} \\
            \kappa_{\text{in}}/2\pi = 3.9(2) \, \text{MHz} \\
            {\color{gray} \kappa_{\text{int, 1}}/2\pi = 39.9(2) \, \text{MHz}}
            \end{array}$ \\
            \hline 
        \end{tabularx}
         \caption{Calibrated dissipation rates for cavity 1.}
        \label{tab:diss_rates_1}
    \end{minipage}
    \hspace{0.5cm}
    \begin{minipage}{0.37\textwidth}
        \centering
        \renewcommand{\arraystretch}{1.5}
        \begin{tabularx}{\textwidth}{|l|X|}
            \hline
            \textbf{Port} & \textbf{Results} \\
            \hline
            Amplifier Port &
            $\begin{array}{l}
            \omega_2/2\pi = 6.0360(6) \, \text{GHz} \\
            \kappa_c/2\pi = 10.1(1) \, \text{MHz} \\
            \kappa_{\text{int,2}}/2\pi = 7.6(1) \, \text{MHz}
            \end{array}$ \\
            \hline
            Reciprocal Port &
            $\begin{array}{l}
            \omega_2/2\pi = 6.0350(6) \, \text{GHz} \\
            \kappa_c/2\pi = 10.0(1) \, \text{MHz} \\
            \kappa_{\text{int, 2}}/2\pi = 7.5(1) \, \text{MHz}
            \end{array}$ \\
            \hline
            Readout Port &
            $\begin{array}{l}
            \omega_2/2\pi = 6.034(5) \, \text{GHz} \\
            \kappa_{\text{out}}/2\pi= 3.6(2) \, \text{MHz} \\
            {\color{gray} \kappa_{\text{int, 2}}/2\pi = 40(2) \, \text{MHz}}
            \end{array}$ \\
            \hline
        \end{tabularx}
        \caption{Calibrated dissipation rates for cavity 2.}
        \label{tab:diss_rates_2}
    \end{minipage}
\end{table}

Note that the dissipation rates and resonance frequencies derived from fitting differ from those used in our models (see ~\supptabref{tab:paramtab}). This discrepancy stems from several factors. Estimating internal dissipation from under-coupled ports such as the drive and readout often leads to imprecise results, hence the uncertainty indicated by graying out these values in ~\supptabref{tab:diss_rates_1} and~\supptabref{tab:diss_rates_2}. Additionally, in the dimer configuration, the resonance frequencies of the cavities decrease by about 7.5 MHz due to loading effects.

The coupling rates obtained from calibration closely match those in the numerical simulations because the stronger coupling at these ports produces a well-defined reflection spectrum, facilitating more accurate parameter estimation. However, the coupling rates used in the simulations are slightly lower than those measured experimentally. This discrepancy could be due to reflections from amplifiers and other components, which might lower the coupling and internal dissipation rates when the system is assembled into the dimer configuration. While precisely quantifying this effect remains a significant challenge, it is important to note that these variations do not fundamentally alter the underlying physics described by our model.

Moreover, the fitting function in Eq.\,\eqref{eq:s11_fit}, originally developed for a 2-port, hanger-type, $\lambda/4$ resonator \cite{S11_fit, S11_fit_transfer_matrix}, is adapted in our experiments by terminating the other ports with 50 $\Omega$ to minimize additional reflections that could distort the reflection spectrum. Given that we use a multi-port device, inherent errors associated with this setup are difficult to quantify solely from fitting uncertainties. Nonetheless, the parameters employed in our numerical simulations closely match those measured experimentally, and any deviations do not significantly impact the physical phenomena that our model describes.

\section{Experimental characterization for the saturable gain non-linearity}

\begin{figure}[htbp!]
\centering
\includegraphics[width=\textwidth]{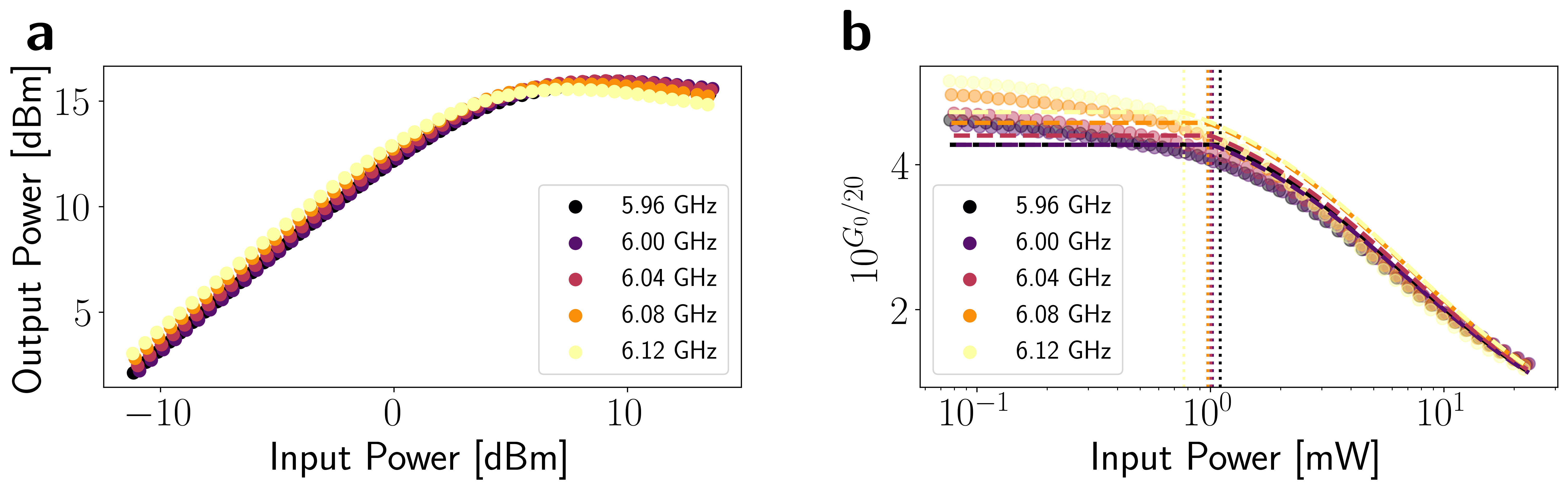}
\caption{\textbf{Characterization of the saturable tunable amplifier across multiple frequencies}. Input–output response of the amplifier and digital attenuator shown in \textbf{a} logarithmic and \textbf{b} linear units, using the conversion $10^{G_0/20}$. Dashed lines represent fits to the piecewise gain model at each measured frequency. Correspondingly, the extracted saturation power, $P_{\text{sat}}$, is marked by a vertical dashed line for each curve.}
\label{fig:saturation-amp-digital-att}
\end{figure}

To validate the applicability of our piecewise gain model, we measured the input–output response of the Minicircuits CMA-83LN+ amplifier and LDA-5018V digital attenuator across five representative frequencies, namely 5.96, 6.00, 6.04, 6.08, and 6.12 GHz, which span the operational range relevant to our experiments. The results are presented in  \suppfigref{fig:saturation-amp-digital-att}, showing the gain as a function of input power in both logarithmic (\textbf{a}) and linear (\textbf{b}) units.

Despite small variations at low input power in  \suppfigref{fig:saturation-amp-digital-att}\textbf{b}, the gain profiles show consistent saturation behavior and converge in the high-power regime, where the amplifier response becomes essentially flat. This confirms that the gain saturation behavior is effectively frequency-independent over the bandwidth of interest. Consequently, the use of a single piecewise function, outlined in Eq. \ref{eq:def_f_lG}, remains valid and sufficient to model the nonlinear response of the active element that we utilize to implement our coupling mechanism in our platform.

Dashed lines in  \suppfigref{fig:saturation-amp-digital-att} correspond to fits using the piecewise saturation function, with extracted values for the saturation power, $P_{\text{sat}}$ and gain curvature parameter, $b_G$. The 1 dB compression point, determined from the gain profile of the amplifier, provides a reliable initial guess for $P_{\text{sat}}$ and is used consistently across all measured frequencies. Based on our measurements, we report $P_{\text{sat}} = 0.995(7)$ mW and $b_G = 7.7(9)$ mW, with uncertainties reflecting the standard deviation across fitted values. These results confirm the consistency of our model and are included to support transparency and reproducibility.

\section{Methodology for Developing the \textit{Hash Map}}
\label{sec:hashmap}

This section provides a detailed overview of the methodology used to develop the \textit{hash map}. In our context, the \textit{hash map} functions as a lookup table that enables the symmetrization of the net gain $(\Delta G)$ across the forward and backward hopping paths. This symmetrization is accomplished by adjusting the digital attenuators $(\Gamma_{1\to 2}$ and $\Gamma_{2\to 1}$) to achieve the targeted $\Delta G$ levels in each path. The process involves mapping the settings $(\Gamma_{1\to 2}$, $\Gamma_{2\to 1}$, $\phi_{\text{exp}}$) to outcomes $(\Delta G, \phi)$, and 
is essential to establish a clear and robust relationship between the experimental ``knobs'' and the desired parameters in operation.

\begin{figure}[!tb]
\centering
\includegraphics[width=.6\textwidth]{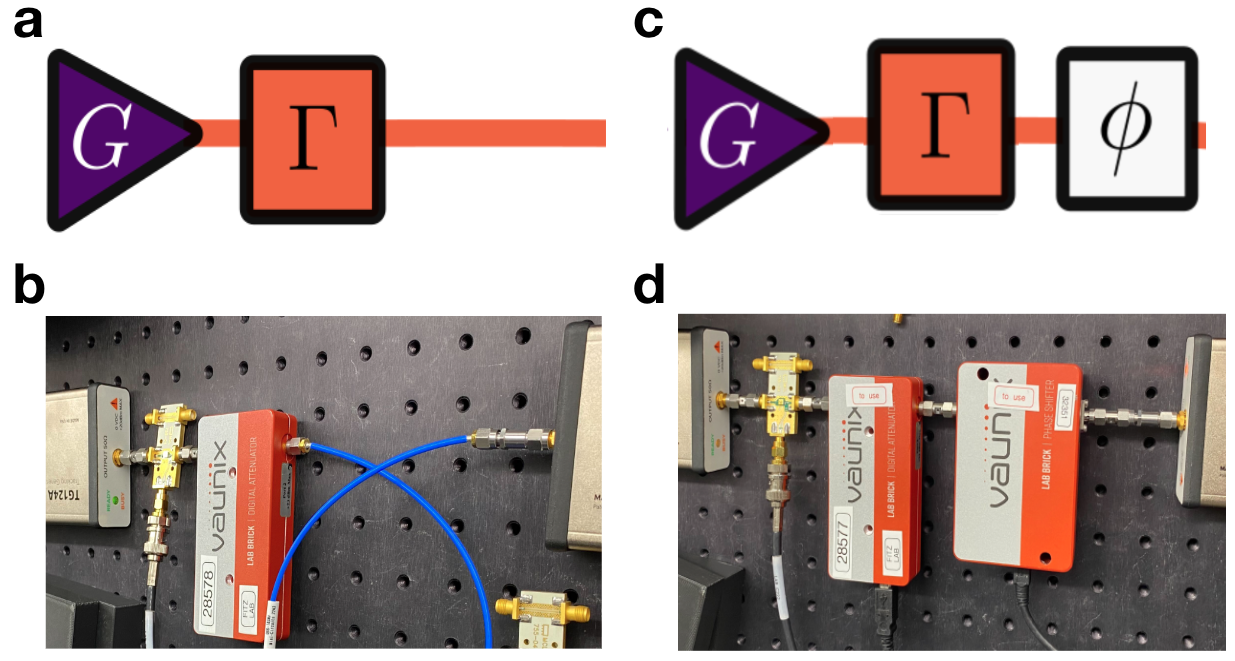}
\caption{\textbf{Schematic for developing the \textit{hash map}:} \textbf{a-b,} Schematics and picture of the experimental setup showing the elements and real devices that facilitate unidirectional hopping in the forward direction ($1 \to 2$), including an amplifier, digital attenuator, and wire between the oscillators. \textbf{c-d} Elements included in the backward direction ($2 \to 1$), comprising an amplifier, digital attenuator, and phase shifter.}
\label{fig:hashmap-schematic}
\vspace*{-3mm}
\end{figure}
\suppfigref{fig:hashmap-schematic} \textbf{a-d} illustrates the experimental setup used to construct the hash map for both the forward (\textbf{a-b}) and backward (\textbf{c-d}) directions, respectively. For the forward path, adjusting $\Delta G$ is straightforward: it involves sampling the recorded transmission $S_{21}$ at various settings of the digital attenuator ($\Gamma_{1\to 2}$). The resulting $S_{21}$ values directly reflect $\Delta G$ for each $\Gamma_{1\to 2}$ and already inherently account for the insertion loss of the components.

\begin{figure}[htbp!]
\centering
\includegraphics[width=.82\textwidth]{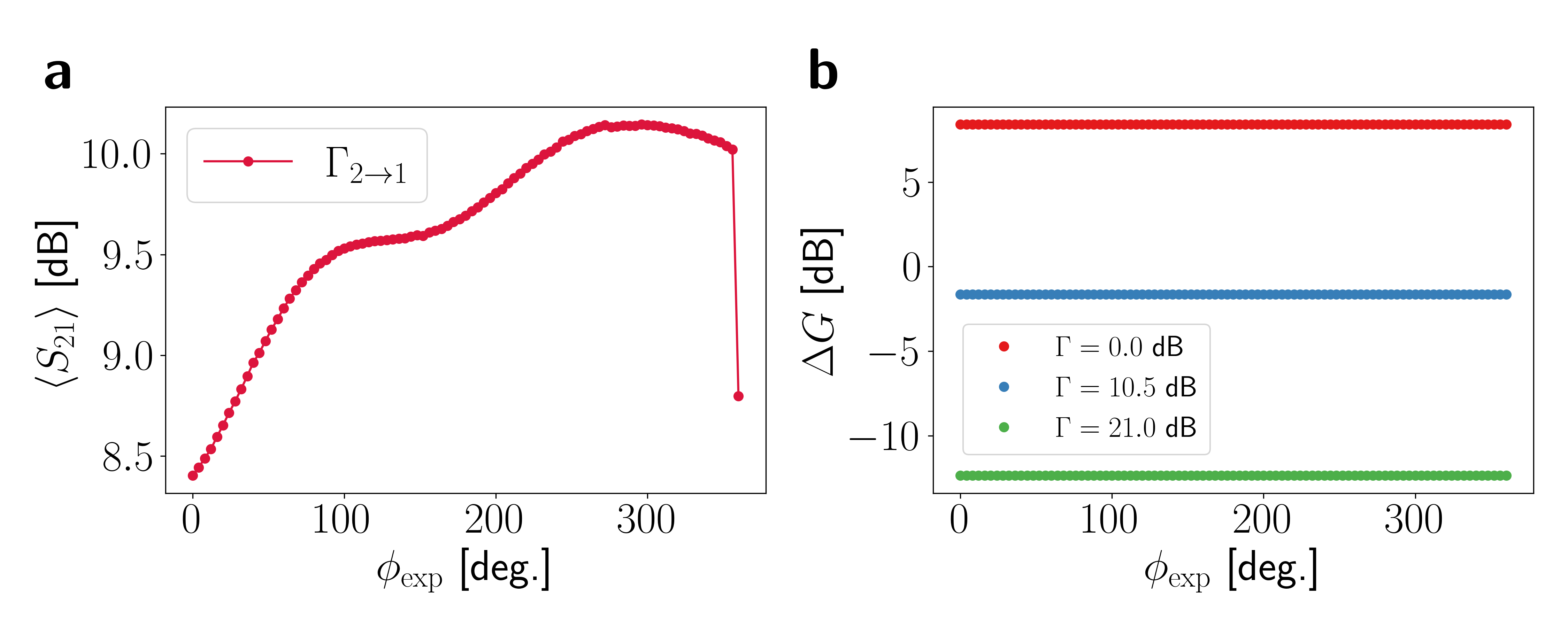}
\vspace*{-5mm}
\caption{\textbf{Calibrating net gain in the backward path (\bm{$2 \to 1$}).} \textbf{a,} Recorded $\langle S_{21} \rangle$ as a function of the set phase, $\phi_{\text{exp}}$, keeping $\Gamma_{2\to1} = 0$ dB. $\langle S_{21} \rangle$ in $\textbf{a}$ represents the average $S_{21}$ over a 200 MHz span around 6 GHz. \textbf{b,} Calibrated profile ensuring a constant $\Delta G$ across $\phi_{\text{exp}}$ for multiple $\Gamma$ values. This calibration is achieved by precisely adjusting the digital attenuator $\Gamma_{2\to1}$ at each $\phi_{\text{exp}}$.}
\vspace*{-5mm}
\label{fig:hashmap-traces}
\end{figure}

However, characterizing $S_{21}$ for the backward path ($2\to1$) is challenging, especially when adjusting the phase settings, $\phi_{\text{exp}}$, on the real device.  \suppfigref{fig:hashmap-traces} \textbf{a} illustrates the measured $S_{21}$ for the backward path across various $\phi_{\text{exp}}$, showing a smooth but non-monotonic trend. This variability could cause significant asymmetries in the hopping coefficients in the dimer configuration, due to inconsistent $S_{21}$ values across different $\phi_{\text{exp}}$. To mitigate this, we normalized the insertion-loss profile of the phase shifter by adjusting the attenuation for each phase setting, effectively mapping $\Gamma_{2\to 1} = \Gamma_{2\to 1}(\phi_{\text{exp}})$ to maintain constant $S_{21}$, hence ensuring a desired $\Delta G$. The successful implementation of this correction is demonstrated in \,\suppfigref{fig:hashmap-traces} \textbf{b}. Furthermore, \suppfigref{fig:hashmap-traces}\textbf{a} clearly shows an outlier where $\phi_{\text{exp}}$ cycles through 360 degrees. We excluded this data point from the phase diagram presented in the main manuscript.

\begin{figure}[htbp!]
\centering
\includegraphics[width=.82\textwidth]{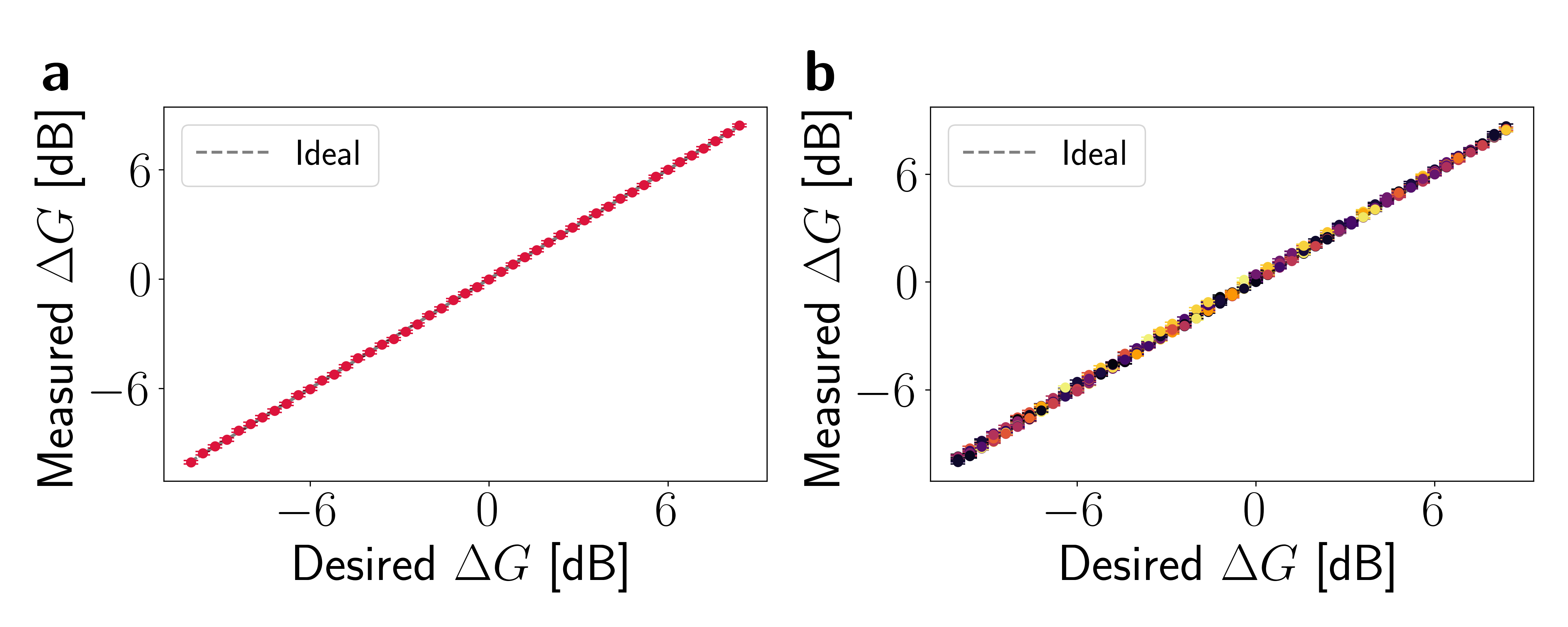}
\vspace*{-5mm}
\caption{\textbf{Measured vs. desired net gain ($\bm{\Delta G}$) levels.} For the \textbf{a,} forward ($1\to2$), and \textbf{b,} backward ($2\to1$) arms. Multiple markers at each desired $\Delta G$ in \textbf{b} correspond to ten randomly sampled $\phi_{\text{exp}}$ values. Both plots are compared against the ideal case, where dashed lines indicate a perfect match between the desired and measured $\Delta G$. Error bars on the markers in \textbf{a-b} represent the standard deviation of transmission ($S_{21}$) measurements over a 200 MHz span.}
\label{fig:hashmap-statistics}
\end{figure}

To verify the robustness of our approach, we conducted statistical tests to confirm that the desired $\Delta G$ values could be experimentally and reliably achieved in operation. \,\suppfigref{fig:hashmap-statistics} \textbf{a-b} display the measured versus desired $\Delta G$ for both the forward ($1\to2$) and backward ($2\to1$) arms. Notably, \,\suppfigref{fig:hashmap-statistics}\textbf{b} demonstrates that $\Delta G$ remains consistent across ten different randomly sampled $\phi_{\text{exp}}$ settings, indicating that our approach is essentially phase-insensitive. Hence, the \textit{hash map} maintains stable $\Delta G$ values across the explored parameter range, highlighting the reliability and precision of our method. In any case, the measured versus desired $\Delta G$ is very close to the ideal case, where both quantities are perfectly aligned.

